\documentclass[]{raa}            

\usepackage{graphicx,times}             
\usepackage{natbib}
\usepackage{amssymb,amsmath}
\usepackage{threeparttable}
\usepackage[T1]{fontenc}

\bibpunct{(}{)}{;}{a}{}{,}

\begin{document}

    \title{Investigation of intergranular bright points from the New Vacuum Solar Telescope
   \footnotetext{$*$ Supported by the National Natural Science Foundation of China.}
   \footnotetext{\textdagger  Corresponding author}
}

   \volnopage{Vol.0 (200x) No.0, 000--000}      
   \setcounter{page}{1}          

   \author{Kai-Fan Ji
        \inst{1}
   \and Jian-ping Xiong
        \inst{1,2,3}
   \and Yong-yuan Xiang
        \inst{4}
    \and Song Feng
        \inst{1,2,3}
   \and Hui Deng
        \inst{1}
   \and Feng Wang
        \inst{1,4}
   \and Yun-Fei Yang
        \inst{1,2,3}
        {\textdagger}
   }

   \institute{Faculty of Information Engineering and Automation / Yunnan Key Laboratory of Computer Technology Application, Kunming University of Science and Technology,
   Kunming 650500, China; {\it yangyf@escience.cn}\\
        \and
             Key Laboratory of Solar Activity, National Astronomical Observatories, Chinese Academy of Sciences, Beijing 100012, China\\
        \and
             Key Laboratory of Modern Astronomy and Astrophysics, Nanjing University, Ministry of Education, Nanjing 210093, China\\
         \and
         Yunnan Observatory, Chinese Academy of Sciences, Yunnan 650011, China\\
   }

   \date{Received~~2015 April 30; accepted~~2015~~July 29}

\abstract{Six high-resolution TiO-band image sequences from the New Vacuum Solar Telescope (NVST) are used to investigate the properties of intergranular bright points (igBPs).
We detect the igBPs using a Laplacian and morphological dilation algorithm (LMD) and track them using a three-dimensional segmentation algorithm automatically, and then investigate the morphologic, photometric and dynamic properties of igBPs, in terms of equivalent diameter, the intensity contrast, lifetime, horizontal velocity, diffusion index, motion range and motion type. The statistical results confirm the previous studies based on G-band or TiO-band igBPs from the other telescopes. It illustrates that the TiO data from the NVST have a stable and reliable quality, which are suitable for studying the igBPs. In addition, our method is feasible to detect and track the igBPs in the TiO data from the NVST.
With the aid of the vector magnetograms obtained from the Solar Dynamics Observatory /Helioseismic and Magnetic Imager, the properties of igBPs are found to be influenced by their embedded magnetic environments strongly. The area coverage, the size and the intensity contrast values of igBPs are generally larger in the regions with higher magnetic flux. However, the dynamics of igBPs, including the horizontal velocity, the diffusion index, the ratio of motion range and the index of motion type are generally larger in the regions with lower magnetic flux. It suggests that the absence of strong magnetic fields in the medium makes it possible for the igBPs to look smaller and weaker, diffuse faster, move faster and further in a straighter path.
}
   \authorrunning{K.-F. Ji, J.-P. Xiong, Y.-Y. Xiang, S. Feng, H. Deng, F. Wang \& Y.-F. Yang}            
   \titlerunning{Intergranular bright points from the New Vacuum Solar Telescope}  

   \maketitle

%
%
\section{Introduction}           
\label{sect:intro}
The New Vacuum Solar Telescope (NVST; \citealt{Wang2013first}; \citealt{Liu2014New}; \citealt{Xu2014Primary}) at Fuxian Solar Observatory of Yunnan Astronomical Observatory in China is designed to observe the Sun with very high spatial and spectral resolution in the wavelength range from 0.3 to 2.5 micron. From October 2012, it mainly provides H$\alpha$ (656.3$\pm$0.025 nm) band data for observing the chromosphere and TiO (705.8$\pm$1 nm) band image for observing the photosphere. Many studies focused on the small-scale structures and fine details in the chromosphere using H$\alpha$ band data have been carried out (\citealt{Yang2014New}; \citealt{Yang2014Fine}; \citealt{Bi2015Partial}; \citealt{Yan2015Formation}; \citealt{Yang2015Magnetic}). However, the studies using the TiO-band data are scarce.

Thought to be the foot-points of the magnetic flux tubes, intergranular bright points (igBPs) are clearly visible in some lines formed in the photosphere, such as G-band, CN band, blue continuum and TiO-band (\citealt{Zakharov2005comparative}; \citealt{Abramenko2010Statistical}). IgBPs show a strong spatial correlation with magnetic flux concentrations and are therefore useful as magnetic proxies, which allow the distribution and dynamics of magnetic features to be studied at a higher spatial resolution than using spectro-polarimetric techniques (\citealt{Keller1992Resolution}, \citealt{Berger2001On}, \citealt{Sanchez2001Thermal}; \citealt{Steiner2001Radiative}; \citealt{Schussler2003Why}; \citealt{Carlsson2004Observational}; \citealt{Shelyag2004Gband}; \citealt{Beck2007Magnetic}; \citealt{Ishikawa2007Relationships}; \citealt{deWijn2008Hinode}).
A theoretical model for formation process of igBPs is called convective collapse model (\citealt{Parker1978Adiabatic}; \citealt{Spruit1979Convective}). The model is suggested that when magnetic field exceeds an equipartition field strength, the plasmas within the magnetic field draft down, resulting in a small scale vertical flux tube that is visible as a bright point. The correlation between the brightness and the field strength is explained by hot-wall mechanism (\citealt{Spruit1976Pressure}; \citealt{Spruit1981size}). Accordingly, the less opaque magnetic flux-tube interior then causes an excess of lateral inflow of radiation into their evacuated interiors, and as a consequence the magnetic elements appear brighter than their surroundings.
The radiative properties of igBPs possibly play an important role in influencing the Earth's climate (\citealt{London1994Observed}; \citealt{Larkin2000Effect}; \citealt{Gray2010Solar}; \citealt{Ermolli2013Recent}; \citealt{Solanki2013Solar}). Moreover, the motions of igBPs can influence the granulation and energy transport process in the lower solar atmosphere (e.g., strong magnetic field can suppress normal convective flows; \citealt{Title1989Statistical}; \citealt{Andic2011Response}). Therefore, the motions can indicate the properties of MHD waves excited at lower solar atmosphere that may contribute to coronal heating, and generate kinetic and Alfv\'{e}n waves and then release energy (\citealt{Roberts1983Wave}; \citealt{Parker1988Nanoflares}; \citealt{Choudhuri1993Implications}; \citealt{deWijn2009On}; \citealt{Jess2009Alfven}; \citealt{Zhao2009Magnetic}; \citealt{Balmaceda2010Evidence}; \citealt{Ji2012Observation}).

In general, G-band (430.5 nm) and TiO-band (705.7 nm) observations are applied to study igBPs in the previous works. G-band observation is regarded as an excellent proxy for studying igBPs because they appear brighter due to the reduced abundance of the CH molecule at higher temperatures (\citealt{Steiner2001Radiative}). Most studies of igBPs have been performed using different G-band observations, such as 48 cm Swedish Vacuum Solar Telescope (SVST; \citealt{Berger1995New}; \citealt{Berger1998Measurements}; \citealt{Berger2001On}), 43.8 cm Dutch Open Telescope (DOT; \citealt{Bovelet2003Dynamics}; \citealt{Nisenson2003Motions}; \citealt{deWijn2005DOT}; \citealt{Feng2013Statistical}; \citealt{Bodnarova2014On}), 1 m Swedish Solar Telescope (SST; \citealt{Sanchez2004Bright}; \citealt{Mostl2006Dynamics}; \citealt{Chitta2012Dynamics}), 76 cm Dunn Solar Telescope (DST; \citealt{Crockett2010Area}; \citealt{Keys2011Velocity}; \citealt{Romano2012Comparative}; \citealt{Keys2013Tracking}; \citealt{Keys2014Dynamic}); seeing-free space-based 50 cm solar optical telescope onboard Hinode (SOT; \citealt{Utz2009size}; \citealt{Utz2010Dynamics}; \citealt{Yang2014Evolution}; \citealt{Yang2015Characterizing}).
Moreover, TiO-band observation is also used to investigate the igBPs, e.g., 1.6 m New Solar Telescope (NST; \citealt{Abramenko2010Statistical}; \citealt{Abramenko2011Turbulent}). The TiO-band images provide an enhanced gradient of intensity around igBPs, which is very beneficial for imaging them. The reason is that the intensity for granules and igBPs is the same as observed in continuum, whereas for dark cool intergranular lanes, the observed intensity is lowered due to absorption in the TiO line since this spectral line is sensitive to temperature (\citealt{Abramenko2010Statistical}).
In spite of different diffraction limits of these telescopes, the statistical properties of igBPs have been agreed as follows: the typical equivalent diameter is about 150\,$\rm km$, which the range of igBP equivalent diameters is from 76 to 400\,$\rm km$; the typical ratio of the maximum intensity of igBP to the mean photospheric intensity is about 1.1, which the range is from 0.8 to 1.8; the mean lifetime is several minutes, which the range is from 2\,--\,20\, $\rm min$; the mean horizontal velocity is 1\,--\,2\,$\rm km $ $\rm s^{-1}$ with the maximum value of 7\,$\rm km $ $\rm s^{-1}$; the mean diffusion index range from 0.7 to 1.8.

In the last decades, the properties of igBPs in quiet Sun (QS) regions and active regions (ARs) have been compared from observation data. For instance,  \citet{Romano2012Comparative} indicated that igBPs in a QS region are brighter and smaller than those in an AR, but \citet{Feng2013Statistical} drew a different conclusion. Most authors agreed that the dynamics of igBPs is attenuated in ARs compared to QS regions (\citealt{Berger1998Measurements}; \citealt{Mostl2006Dynamics}; \citealt{Keys2011Velocity}). \citet{Keys2014Dynamic} defined one QS sub-region and two active sub-regions in a same FOV judging by their mean Line-of-sight magnetic flux densities. They proposed that the size of igBPs in the QS region is smaller and the horizontal velocity is greater. Besides observations, \citet{Crockett2010Area} utilized mean magnetic fields of 100\,$\rm G$, 200\,$\rm G$, and 300\,$\rm G$ in the magnetohydrodynamic simulations and compared the igBP size with the DST observations. They suggested that the igBP size does not depend on the embedded magnetic environments significantly. \citet{Criscuoli2013Comparison} also analyzed results from simulations characterized by different amounts of average magnetic flux. They indicated that the igBPs decrease in the intensity contrast with increasing environmental magnetic flux. However, few works have been carried out from observations to investigate the differences of the properties of igBPs embedded in varying magnetic fluxes.

The aim of this paper is to investigate the igBPs using the TiO-band data observed from the NVST. Six high-resolution image sequences are selected span from 2012 to 2014, which have different heliocentric angles and magnetic fluxes. We detect and track the igBPs automatically, and investigate the morphologic, photometric and dynamic properties of igBPs, in terms of equivalent diameter, the intensity contrast, lifetime, horizontal velocity, diffusion index, motion range and motion type.
In addition, we investigate the relation between the igBP properties and the amount of magnetic flux of the region they are embedded in. The layout of the paper is as follows: observations and data reductions are described in Section~\ref{sect:Obs}. The method is detailed in Section~\ref{sect:method}. In Section~\ref{sect:result}, the statistics of igBPs in different magnetized environments are presented and discussed, followed by the conclusion in Section~\ref{sect:conclusion}.

\section{Observations and data reductions}
\label{sect:Obs}

The NVST is a vacuum solar telescope with a 985 mm clear aperture, which is designed to observe multi-wavelength high spatial and spectral resolution data. It uses a broadband TiO filter centered at a wavelength of 705.8 nm for observing the photosphere. The team provides the level 1$^{+}$ data, which are processed by frame selection (lucky imaging; \citealt{Tubbs2004Lucky}), and reconstructed by speckle masking (\citealt{Lohmann1983Speckle}) or iterative shift and add (\citealt{Zhou1998Electronic}).
The reconstructed images under the best seeing conditions can almost have a high angular resolution near the diffraction limit of the NVST (105\,$\rm km$) even without the adaptive optics system (\citealt{Liu2014New}). We selected six high-resolution image sequences under the best seeing conditions without adaptive optics system from October 2012 to October 2014. And then, six sub-regions of equal dimensions ( 20$''\times$20$''$) were extracted from the six data sets, respectively. The observation parameters are listed in Table~\ref{Tab:tab1}. Note that, the pixel sizes of the data sets after 2014 are different. It is because the NVST team changed their optical system on 19 May 2014 and resulted in the difference.
The images in each sequence were aligned by a sub-pixel level image registration procedure (\citealt{Feng2012A}; \citealt{Yang2015Characterizing}). The projection effects of the data sets that are away from the solar disk center were corrected according to the heliocentric longitude and latitude of each pixel.

\begin{table}[htb]
\begin{center}
\caption[]{The Parameters of Six Data Sets}\label{Tab:tab1}
\begin{threeparttable}

  \begin{tabular}{lccccccc}
  \hline\noalign{\smallskip}
  Data set & Date & Time interval (UT) & Center of the FOV ($''$) &  Pixel size ($''$) & Cadence(s) & $B$ (G)\tnote{1}\\
  \hline\noalign{\smallskip}
1   &2013-05-21	    &06:14:05--07:30:50 &(-232, 358)   &0.041  &55  &94\\
2   &2013-06-12	    &07:45:34--09:07:29	&(-173, -170)  &0.041  &55 &127\\
3   &2014-10-03 	&04:35:50--05:25:32	&(25, -108)    &0.052  &30  &143\\
4   &2013-07-15 	&07:29:09--08:22:22	&(285, -313)   &0.041  &40 &169\\
5   &2014-09-13 	&02:29:23--03:19:06	&(-445, 126)   &0.052  &30 &191\\
6   &2012-10-29     &06:03:37--06:43:25 &(-436, -279)  &0.041  &40 &229\\

  \noalign{\smallskip}\hline
\end{tabular}
 \begin{tablenotes}
        \footnotesize
        \item[1] $B$: Mean magnetic flux density.
 \end{tablenotes}
\end{threeparttable}
\end{center}
\end{table}

\begin{figure}
\includegraphics[width=0.41\textwidth]{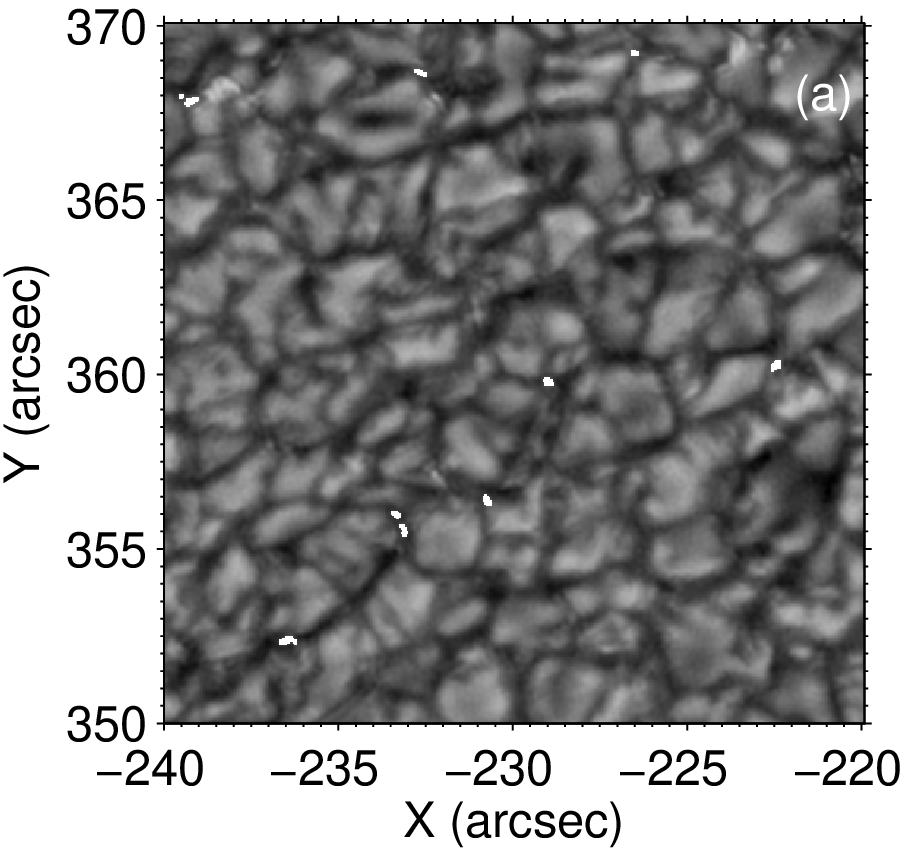}
\includegraphics[width=0.53\textwidth]{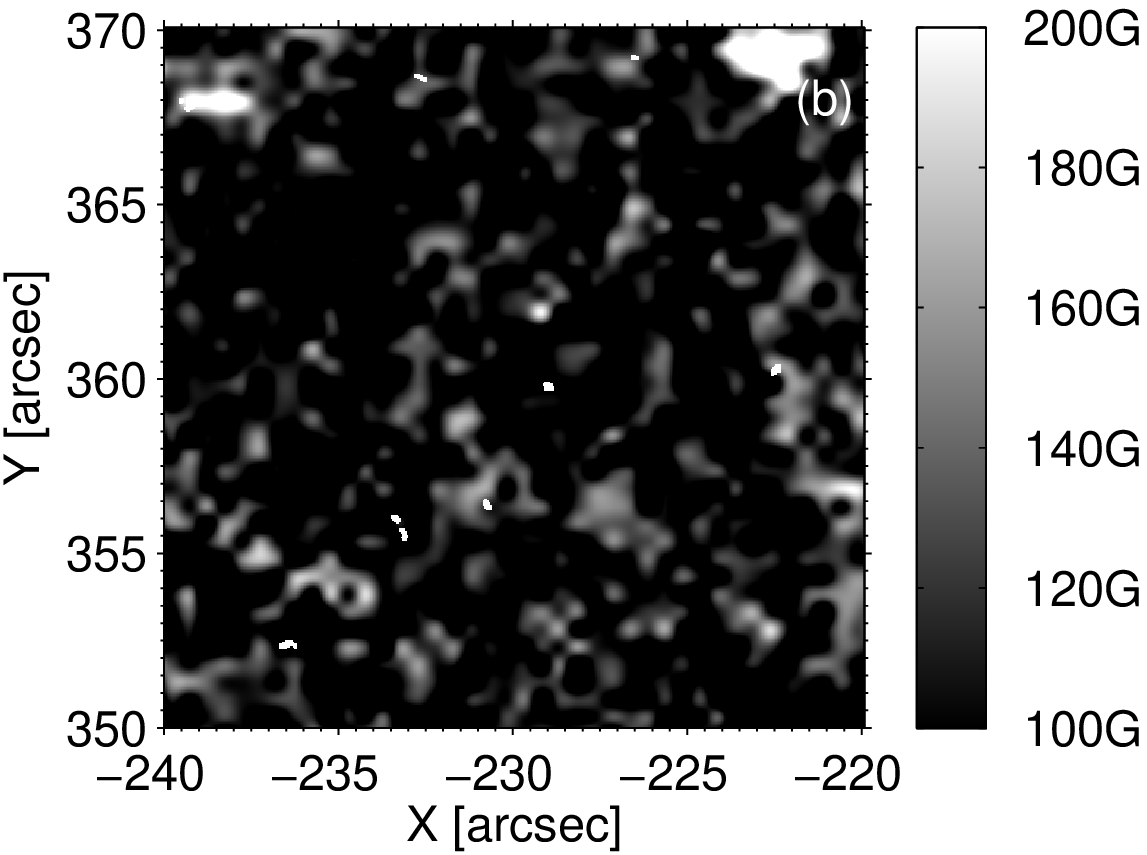}
\caption{(a): one TiO-band image was recorded at 06:14:05 UT on 2013 May 21 from the NVST, in which the detected igBPs are marked with white. (b): the co-spatial HMI vector magnetogram, which the mean magnetic flux density is 94\,$\rm G$ and the maximum value is 321\,$\rm G$. The detected igBPs are also marked with white. The color bar in (c) indicates the magnetic flux density, saturated at [100, 200]\,$\rm G$ to assist visualisation of the field complexities. Axes are in units of arcseconds from the solar disk center.}
    \label{Fig:fig1}
\end{figure}

\begin{figure}
\includegraphics[width=0.41\textwidth]{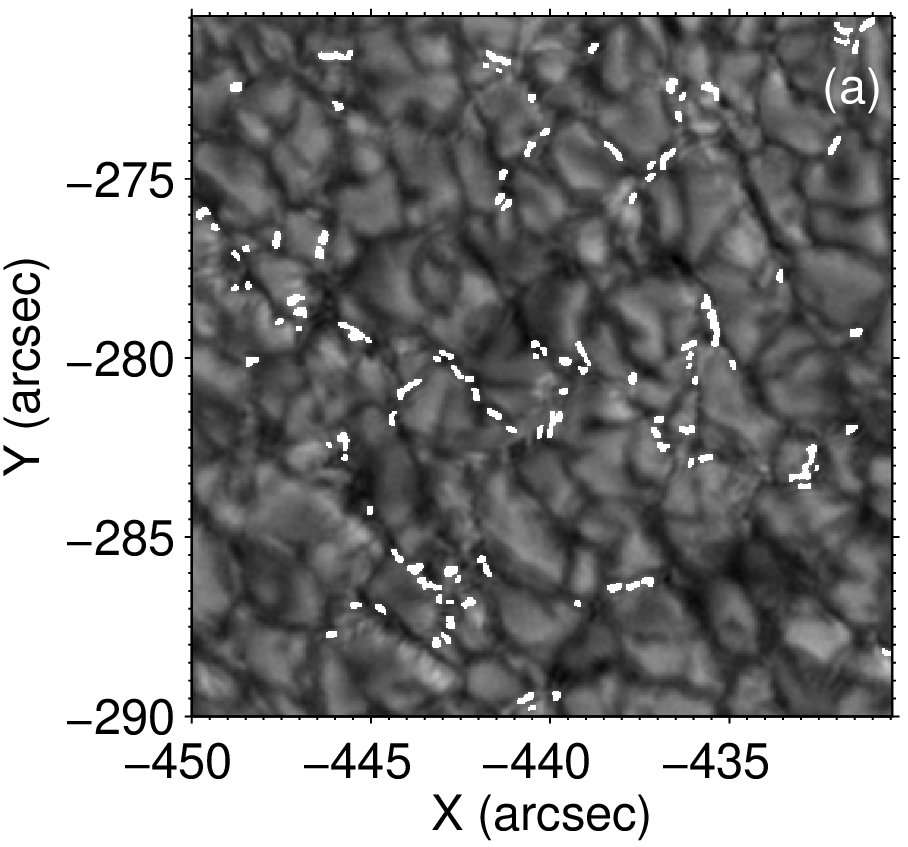}
\includegraphics[width=0.53\textwidth]{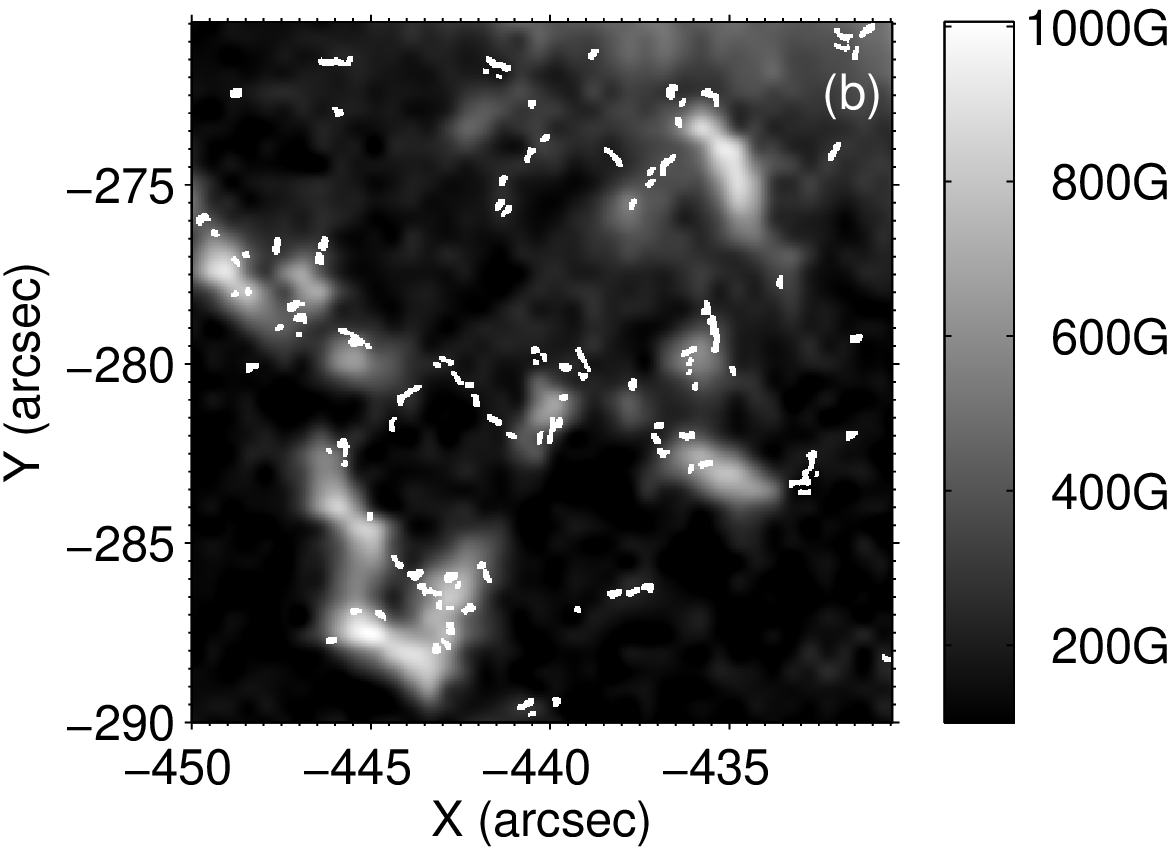}
\caption{(a): one TiO-band image was recorded at 06:03:37 UT on 2012 October 29 from the NVST, in which the detected igBPs are marked with white. (b): the co-spatial HMI vector magnetogram, which the mean magnetic flux density is 228\,$\rm G$ and the maximum value is 1006\,$\rm G$. The detected igBPs are also marked with white. The color bar in (c) indicates the magnetic flux density, saturated at [100, 1006]\,$\rm G$. Axes are in units of arcseconds from the solar disk center.}
    \label{Fig:fig2}
\end{figure}

In order to investigate and compare the properties of igBPs embedded in regions characterized by different average values of magnetic flux, we used the vector magnetograms observed with the Helioseismic and Magnetic Imager (HMI; \citealt{Schou2012Polarization}) on-board the Solar Dynamics Observatory (SDO; \citealt{Pesnell2012Solar}). The HMI data were processed with the standard hmi\_prep routine in SolarSoftware. With the aid of the HMI continuum images, the TiO-band images were co-aligned with the vector magnetograms by the sub-pixel registration algorithm (\citealt{Feng2012A}; \citealt{Yang2015Characterizing}) and the subfields were chosen from the vector magnetograms. Subsequently, the co-aligned sub-magnetograms during the observed time interval of each data set were averaged to improve the sensitivity of the mean magnetic flux density. Data sets are listed in Table~\ref{Tab:tab1} in the order of the corresponding mean magnetic flux density, $B$. Figure~\ref{Fig:fig1} and Figure~\ref{Fig:fig2} show two data sets with the lowest and the highest $B$, which were recorded on 2013 May 21 and 2012 October 29, respectively.

\section{Methods}
\label{sect:method}
A Laplacian and morphological dilation algorithm (LMD; \citealt{Feng2013Statistical}) was used to detect the igBPs in each image. The algorithm consists of three main steps: firstly, the smoothed TiO image is convolved with a Laplacian operation to yield a Laplacian image; secondly, the Laplacian image is applied a threshold $\mu$+3$\sigma$ to produce a binary image, where the $\mu$ and $\sigma$ are the mean value and the standard deviation of the Laplacian image, respectively; finally, the igBPs are filtered by selecting the features whose lengths of the edges are 70 percent inside the intergranular lanes from the binary image. Two samples of the identified igBPs are marked with white in Figure~\ref{Fig:fig1}(b) and Figure~\ref{Fig:fig2}(b).

\begin{figure}
\centering
\includegraphics[width=0.45\textwidth]{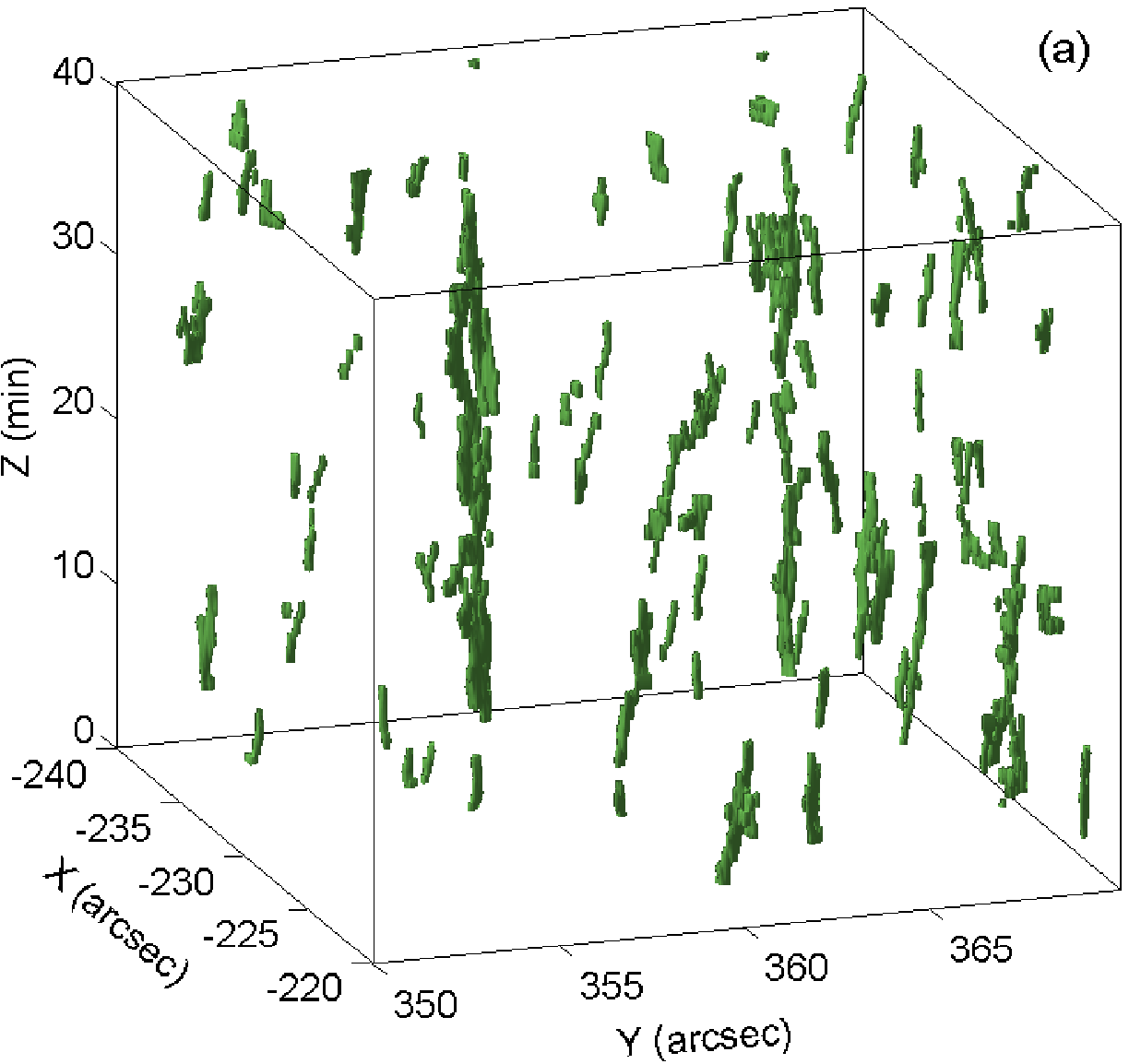}
\hspace {0.3 cm}
\includegraphics[width=0.45\textwidth]{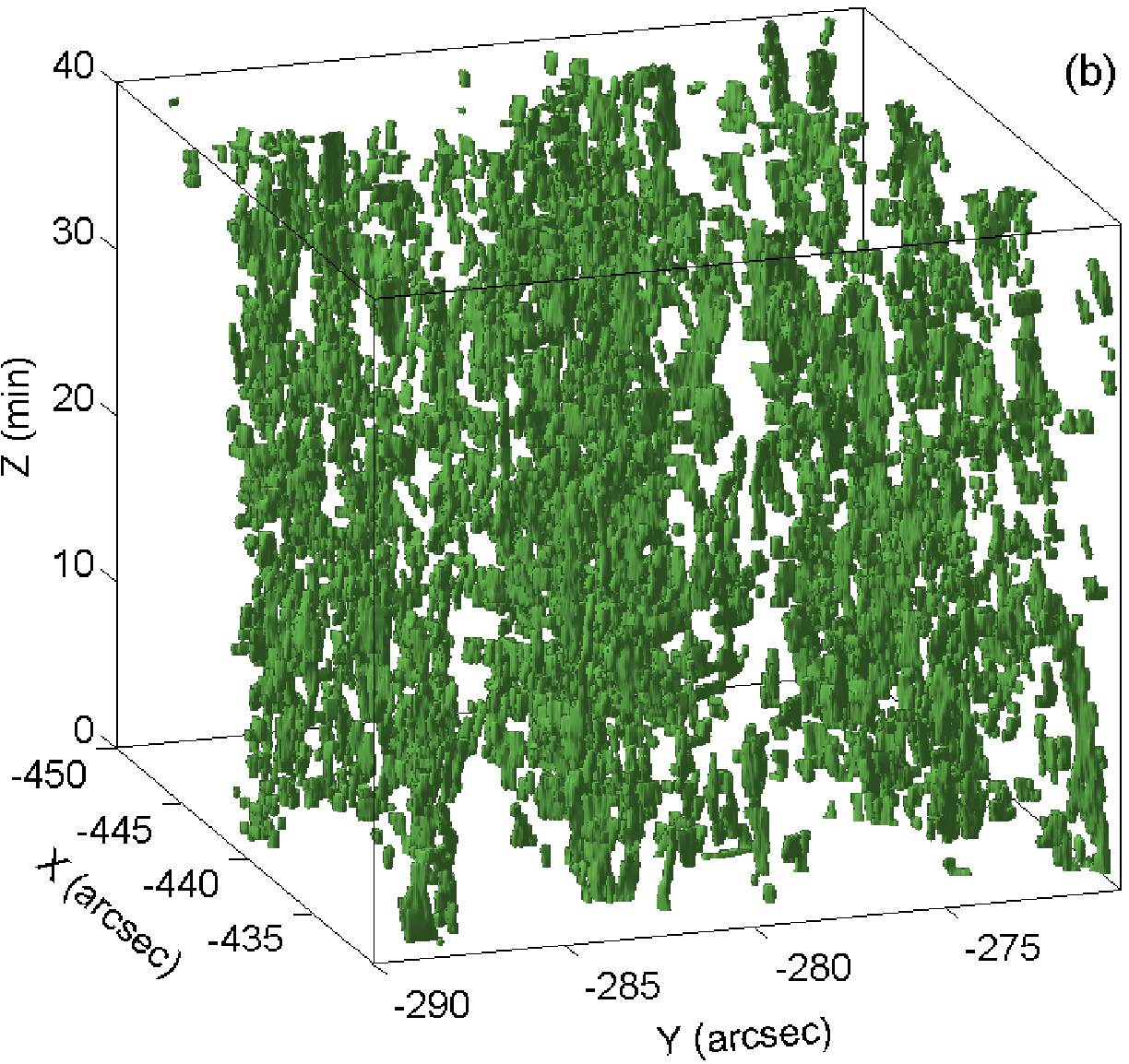}
\caption{Left panel: one segment of the 3D space-time cube whose size is 20$''$$\times$20$''\times$40 min of data set 1. The $x$ and $y$ axes correspond to those in Figure~\ref{Fig:fig1}(a), and the $z$ axis represents time interval from 06:14:05 UT to 06:54:05 UT. Right panel: one segment of the 3D space-time cube igBPs whose size is 20$''$$\times$20$''\times$40 min of data set 6. The $x$ and $y$ axes correspond those in Figure~\ref{Fig:fig2}(a), and the $t$ axis represents time interval from 06:03:37 UT to 06:43:37 UT. The evolution of an igBP presents a 3D structure in the 3D space-time cube.}
    \label{Fig:fig3}
\end{figure}

After detecting the igBPs in each image, a three-dimensional (3D) segmentation algorithm (\citealt{Yang2014Evolution}) was employed to track the evolution of igBPs in the image sequence. The image sequence is regarded as a 3D space-time cube ($x$, $y$, $z$), which the $x$ and $y$ axes are the two dimensional image coordinates, and the $z$ axis represents the frame index of the image sequence. Based on a 26-adjacent technique (\citealt{Yang2013Automatic}; \citealt{Yang2014Evolution}), the evolution of an igBP presents a 3D structure in a 3D space-time cube. Figure~\ref{Fig:fig3} shows two samples of the 3D space-time cubes of data set 1 and 6.

The isolated igBPs that do not merge or split during their lifetimes are selected here because their size, intensity, lifetime and velocity are clearly defined. We also discarded the isolated igBPs with incomplete life cycles or lifetimes are less than 100\,$\rm s$. The rest igBPs (isolated igBPs with complete life cycles) are focused in this study. Table~\ref{Tab:tab2} list the numbers of total igBPs, the non-isolated igBPs, the isolated igBPs with incomplete life cycles and the isolated igBPs with complete life cycles.

\begin{table}
\begin{center}
\caption[]{The numbers of the IgBPs of the Six Data Sets}\label{Tab:tab2}
  \begin{tabular}{lcccccc}
  \hline\noalign{\smallskip}
  Data set & 1 & 2 & 3 & 4 & 5 & 6 \\
  \hline\noalign{\smallskip}
  Total igBPs                                   & 170   &506    &1190   &991    &2071   &1177\\
  Non-isolated igBPs                            & 59    &223    & 506   & 218   & 464   &308\\
  Isolated igBPs with incomplete life cycles    & 9	    &13	    &39	    &66	    &79	    &77\\
  Isolated igBPs with complete life cycles      & 102	&270	&645	&707	&1528	&792\\
  Non-stationary isolated igBPs with complete life cycles    & 75     &190    &440    &290    &625    &337\\

  \noalign{\smallskip}\hline
\end{tabular}
\end{center}
\end{table}

\section{Result and discussion}
\label{sect:result}

\subsection{Statistical properties of igBPs}
\label{sect:statistical}

The area coverage is defined as the percentage of the fractional area occupied by the igBPs. The values of the six data sets are listed in Table~\ref{Tab:tab3}. They range from 0.2\% to 2\%, which are consistent with the most previous studies of 0.5\%--3\% (see Table~\ref{Tab:tab4}).
After that, we calculated the properties of igBPs, in terms of equivalent diameter, intensity contrast, lifetime, horizontal velocity, diffusion index, motion range and motion type. The equivalent diameter is calculated with $\sqrt{4A/\pi}$, where $A$ denotes the area of an igBP. The intensity contrast is defined as the ratio of the peak intensity of an igBP to the average intensity of a quiet sub-region in the FOV. The lifetime is determined by the number of frames over which its corresponding 3D structure. The horizontal velocity is calculated by the displacement of the two centroids between successive frames of an igBP, which the centroid is the arithmetic mean position of all the pixels in the shape of an igBP in a frame. The other properties will be detailed below. Table~\ref{Tab:tab3} list the mean values, the standard deviations, and the ranges of all igBP properties for each data set.

\begin{table}
\begin{center}
\caption[]{The Properties of the IgBPs of the Six Data Sets}\label{Tab:tab3}
  \begin{tabular}{lcccccc}
  \hline\noalign{\smallskip}
  Data set & 1 & 2 & 3 & 4 & 5 & 6 \\
  \hline\noalign{\smallskip}
Area coverage              &0.20\%	&0.99\% &1.55\%   &1.53\%  &1.75\%  &1.99\%\\
Equivalent diameter ($\rm km$)        &181$\pm$22 	&168$\pm$29 	&178$\pm$29 	&195$\pm$36 	&184$\pm$38     &194$\pm$36\\
\quad\quad\quad [\emph{min,max}]      &[\emph{111, 245}]   &[\emph{103, 402}]   &[\emph{109, 440}]   &[\emph{106, 445}]   &[\emph{122, 447}]   &[\emph{112, 473}] \\
Intensity contrast              &0.99$\pm$0.04  &1.01$\pm$0.04  &1.03$\pm$0.04  &1.05$\pm$0.06  &1.05$\pm$0.04  &1.06$\pm$0.05\\
\quad\quad\quad [\emph{min,max}] &[\emph{0.91, 1.12}]   &[\emph{0.90, 1.19}]   &[\emph{0.90, 1.31}]   &[\emph{0.92, 1.30}]   &[\emph{0.92, 1.24}]   &[\emph{0.89, 1.28}]\\
Lifetime ($\rm sec$)            &104$\pm$104     &133$\pm$133     &114$\pm$114     &141$\pm$141     &121$\pm$121     &124$\pm$124\\
\quad\quad\quad [\emph{min,max}] &[\emph{103, 582}]   &[\emph{102, 826}]   &[\emph{120, 723}]   &[\emph{119, 735}]   &[\emph{120, 572}]   &[\emph{114, 580}]\\
Velocity ($\rm km$ $\rm s^{-1}$) &1.35$\pm$0.71  &1.23$\pm$0.64  &1.06$\pm$0.55  &1.04$\pm$0.54  &1.06$\pm$0.55  &1.05$\pm$0.55\\
\quad\quad\quad [\emph{min,max}] &[\emph{0.01, 5.27}]   &[\emph{0, 6.80}]   &[\emph{0.08, 5.43}]   &[\emph{0.02, 5.32}]   &[\emph{0.06, 5.21}]   &[\emph{0.03, 5.75}]\\
Diffusion index                 &1.31$\pm$0.65  &1.21$\pm$0.78  &0.91$\pm$0.43  &1.05$\pm$0.67  &0.86$\pm$0.49  &0.93$\pm$0.77\\
\quad\quad\quad [\emph{min,max}] &[\emph{-3.64, 3.93}]   &[\emph{-4.91, 5.39}]   &[\emph{-4.17, 4.28}]   &[\emph{-5.21, 6.51}]   &[\emph{-7.00, 4.21}]   &[\emph{-5.70, 4.43}]\\
Ratio of motion range           &1.30$\pm$0.80      &1.18$\pm$0.76  &1.11$\pm$0.77  &1.02$\pm$0.62  &0.96$\pm$0.67  &1.03$\pm$0.69\\
\quad\quad\quad [\emph{min,max}] &[\emph{0.31, 5.06}]   &[\emph{0.15, 6.39}]   &[\emph{0.04, 6.28}]   &[\emph{0.17, 5.42}]   &[\emph{0.13, 4.73}]   &[\emph{0.13, 4.79}]\\
Motion type                     &0.69$\pm$0.69  &0.69$\pm$0.69  &0.58$\pm$0.58  &0.59$\pm$0.59  &0.59$\pm$0.59  &0.62$\pm$0.62\\
\quad\quad\quad [\emph{min,max}] &[\emph{0.08, 0.99}]   &[\emph{0.04, 1.00}]   &[\emph{0.23, 0.99}]   &[\emph{0.03, 0.99}]   &[\emph{0, 1.00}]   &[\emph{0.04, 0.98}]\\

  \noalign{\smallskip}\hline
\end{tabular}
\end{center}
\end{table}

\begin{table}
\begin{center}
\caption[]{The Area coverage of the IgBPs of Some Previous Studies}\label{Tab:tab4}
  \begin{tabular}{lcccccc}
  \hline\noalign{\smallskip}
  Reference & Telescope & Region    &Magnetic & Spatial  & Temporal   & Area coverage\\
            &           &           & fluxes        &resolution & resolution ($\rm s$)&\\
    \hline\noalign{\smallskip}
  \citet{Berger1995New}             & SVST  & AR     &          & 0.$^{''}$083      &    &1.8\%\\
  \citet{Feng2013Statistical}       & DOT   & QS    &           & 0.$^{''}$071      & 30    & 0.5\%  \\
  \citet{Feng2013Statistical}       & DOT   & AR    &           & 0.$^{''}$071      & 30    & 1.4\%  \\
  \citet{Sanchez2004Bright}         & SST   & QS    & 0.9 G     & 0.$^{''}$041      & 15    & 0.5\%$-$0.7\% \\
  \citet{Sanchez2010Magnetic}       & SST   & QS    & 29 G      & 0.$^{''}$041      & 15    & 0.9\%$-$2.2\%  \\
  \citet{Romano2012Comparative}     & SST   & QS    &           & 0.$^{''}$041      & 89    & 0.8\%  \\
  \citet{Romano2012Comparative}     & SST   & AR    &           & 0.$^{''}$041      & 20    & 1.9\%  \\

  \noalign{\smallskip}\hline
\end{tabular}
\end{center}
\end{table}

\begin{figure}[!htp]
\centering
        \includegraphics[width=0.4\textwidth]{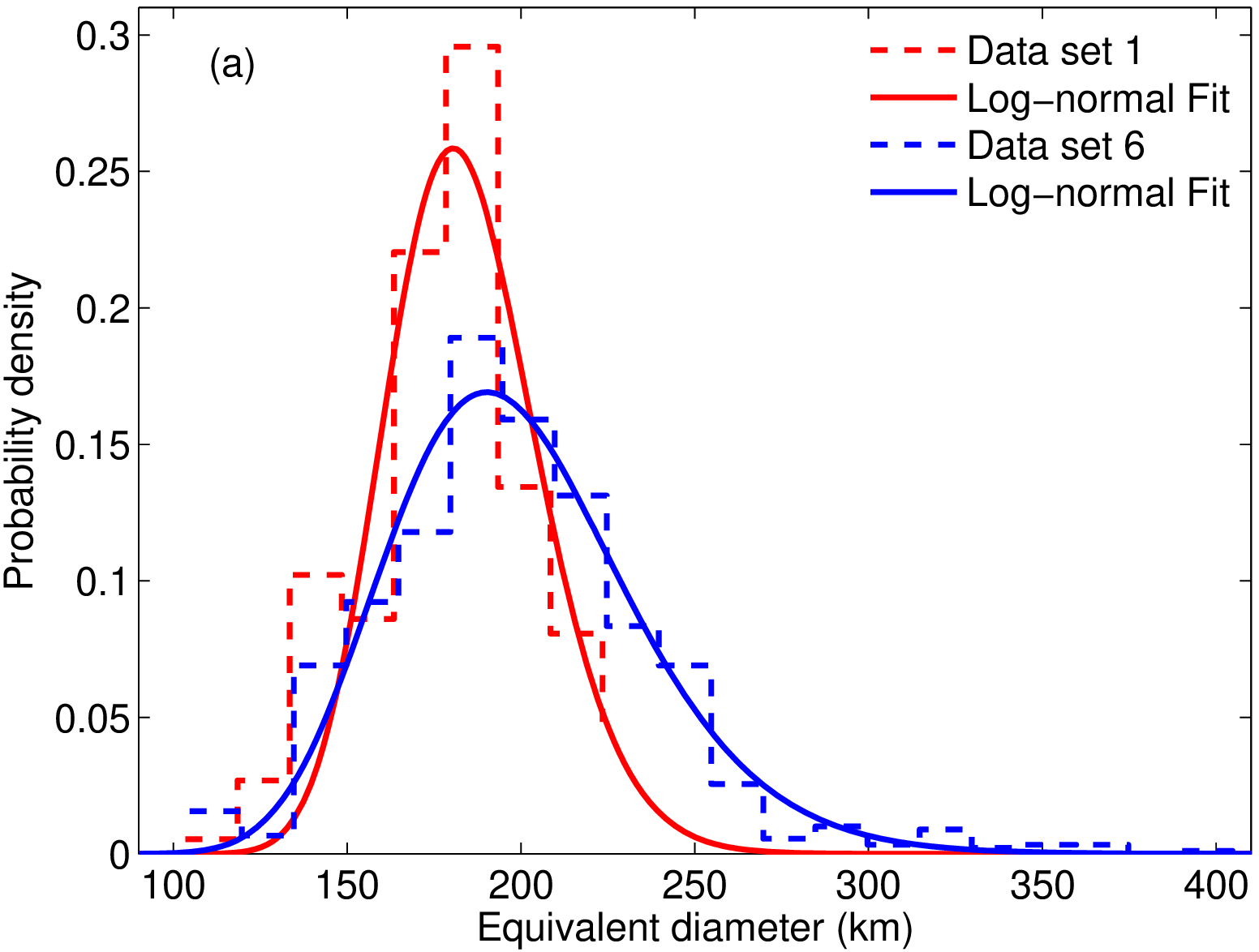}
        \includegraphics[width=0.4\textwidth]{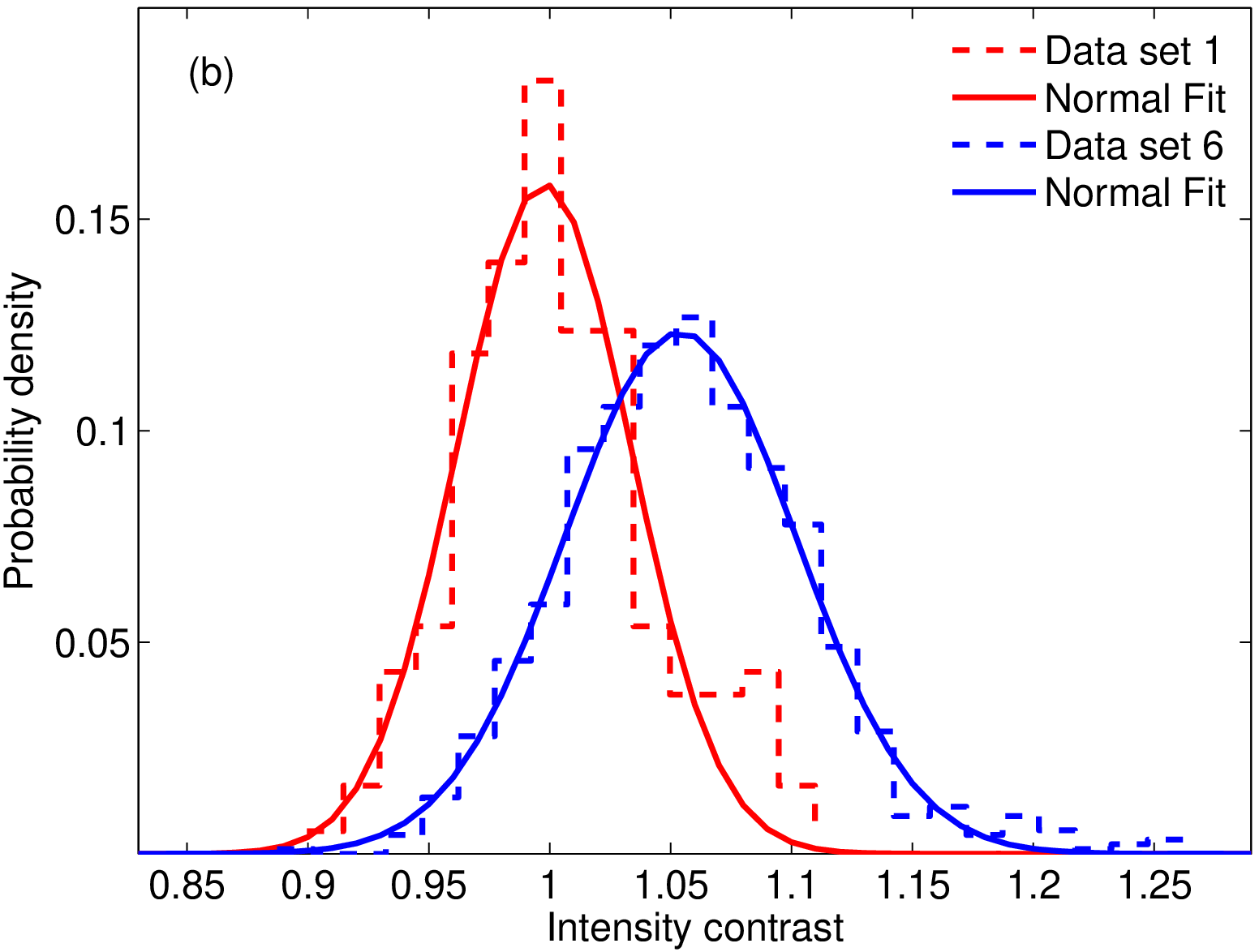}
        \includegraphics[width=0.4\textwidth]{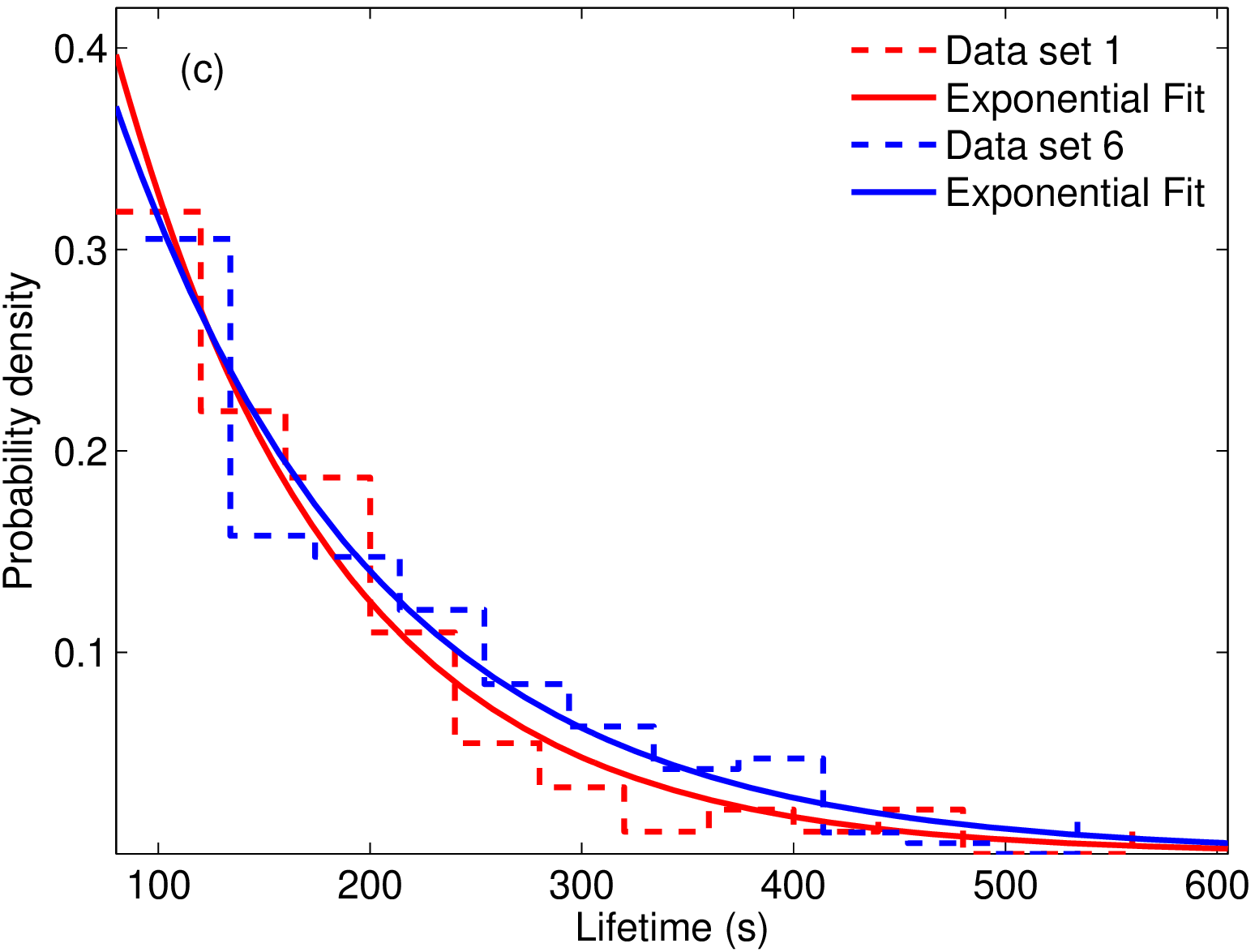}
        \includegraphics[width=0.4\textwidth]{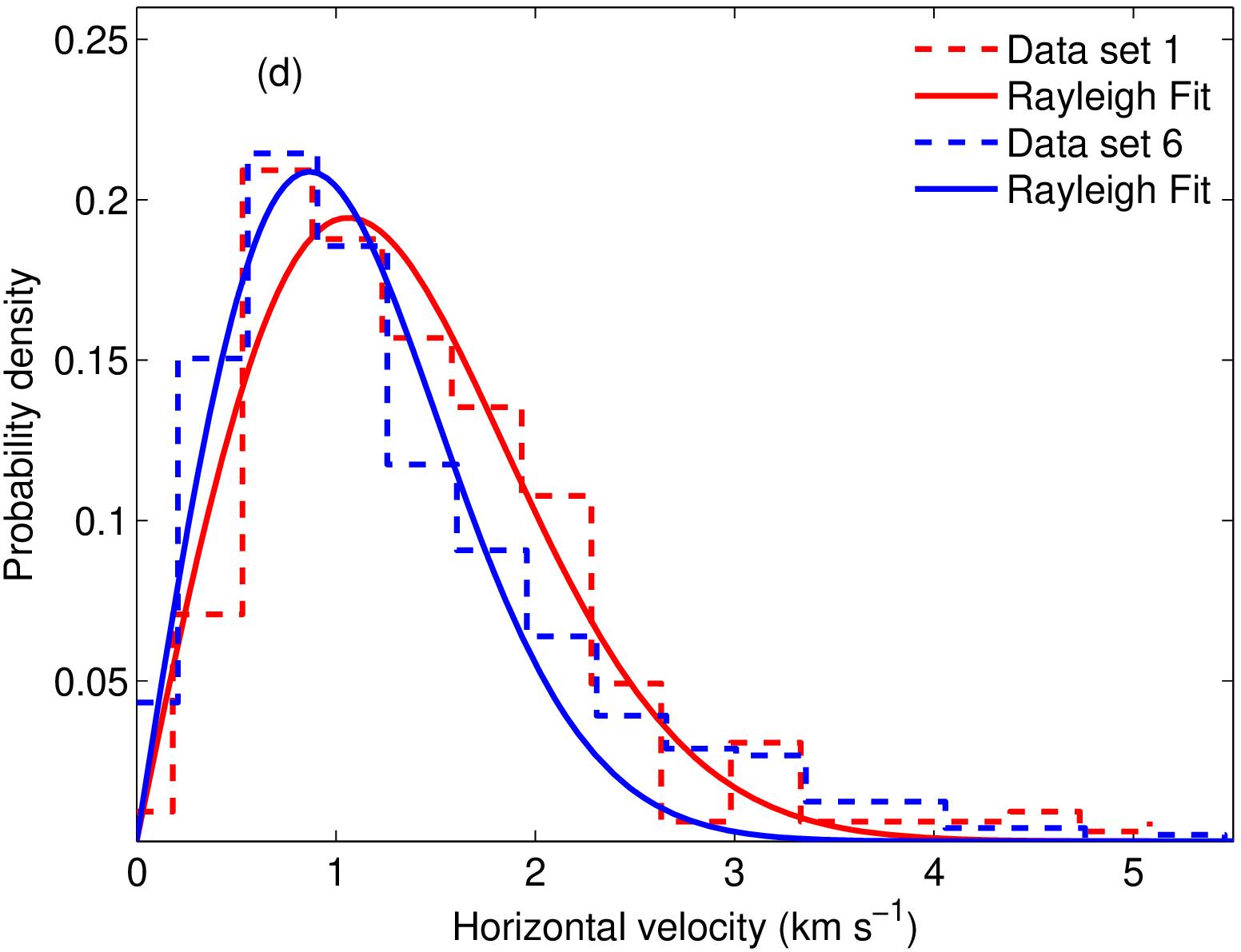}
        \includegraphics[width=0.4\textwidth]{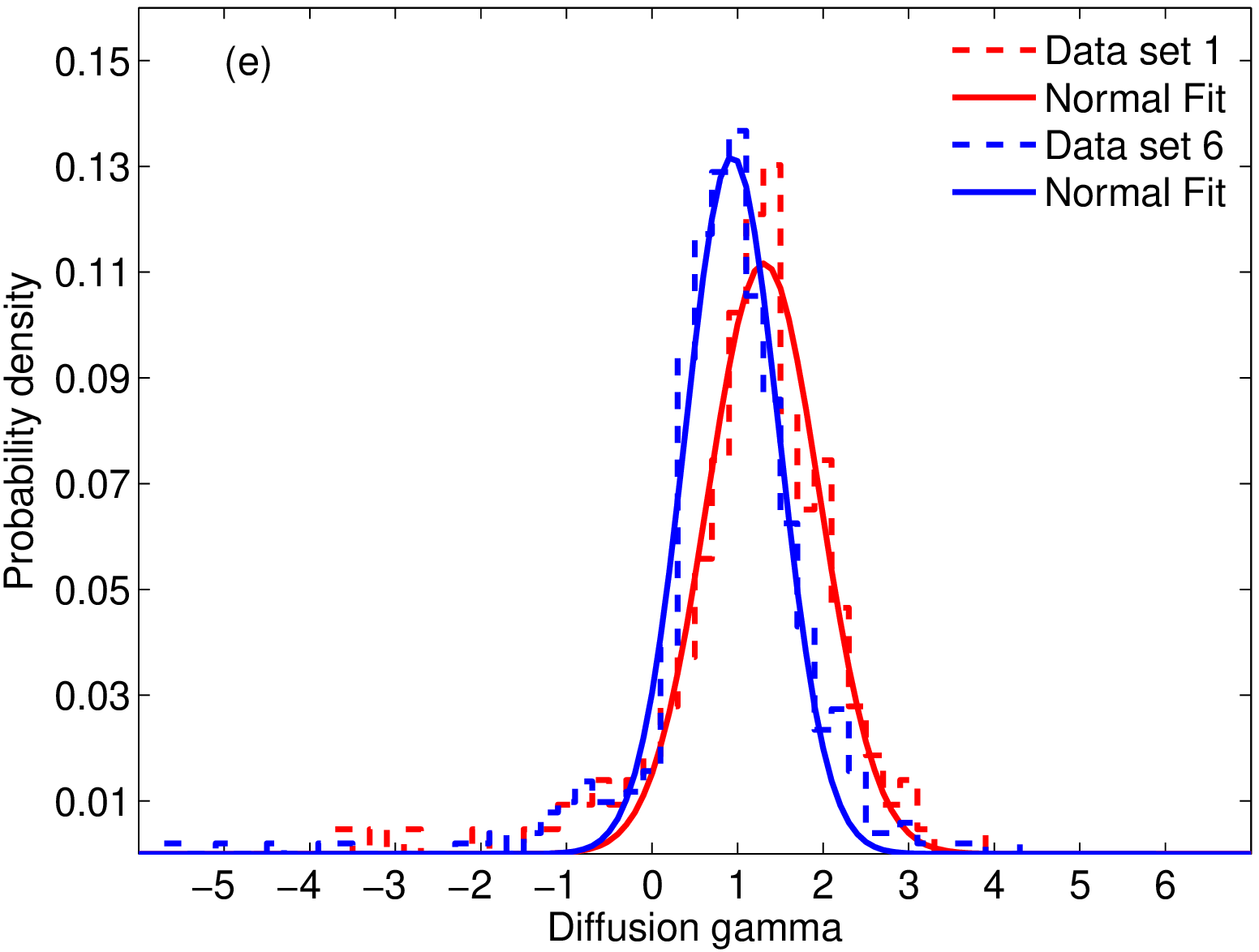}
        \includegraphics[width=0.4\textwidth]{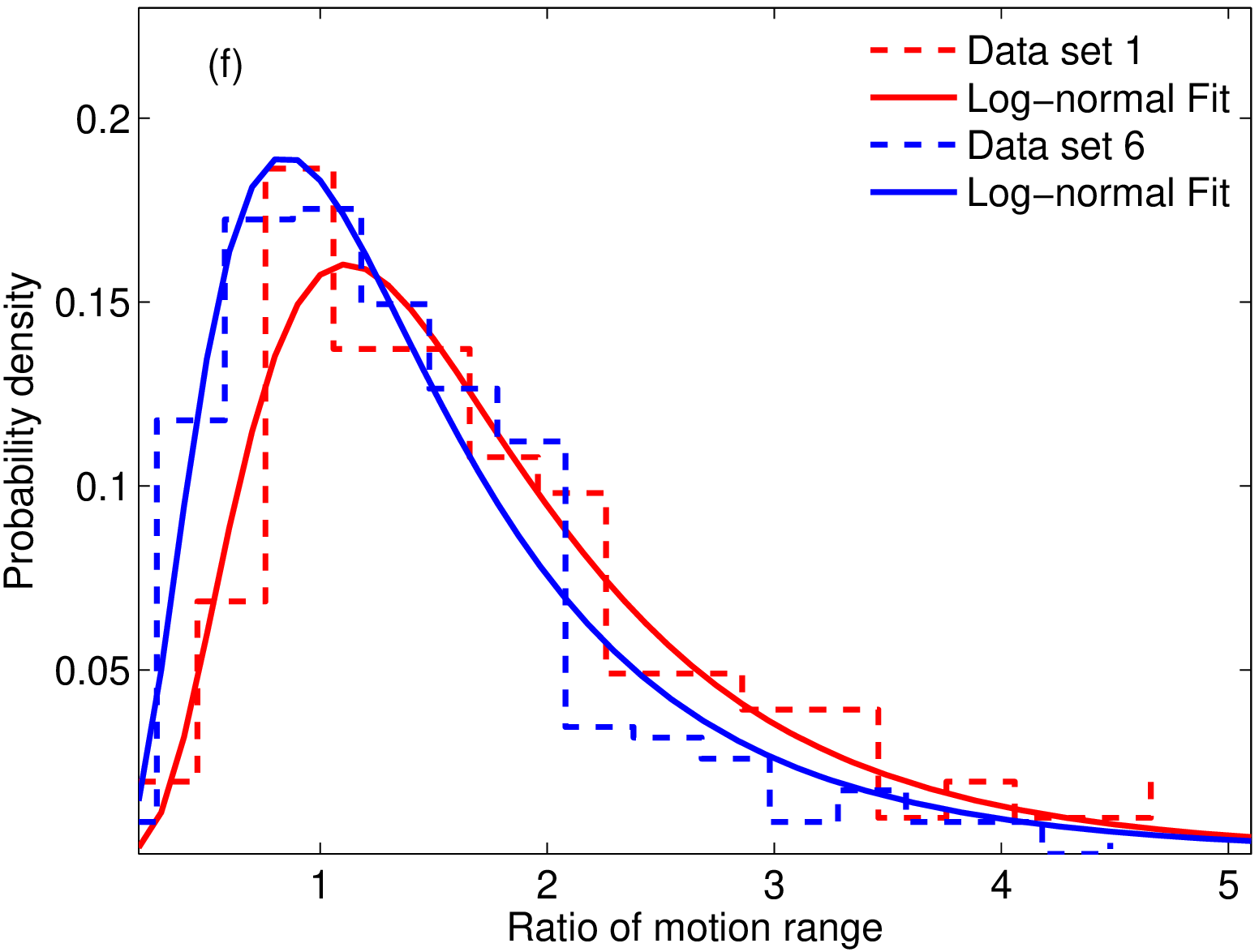}
        \includegraphics[width=0.4\textwidth]{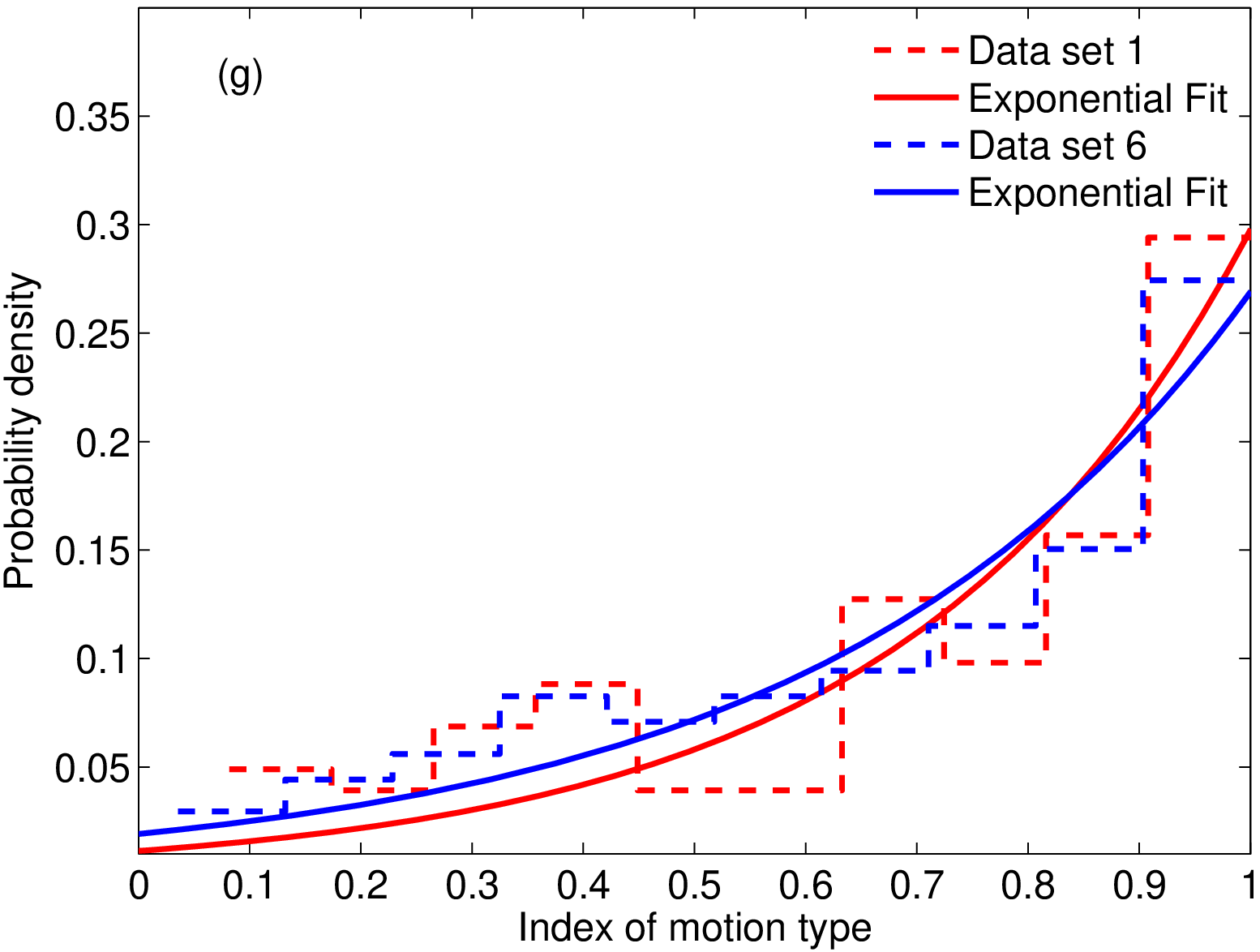}
        \caption{The histograms and the fitted curves of igBP properties of data set 1 and 6. The red lines represent the histograms (dashed) and the distribution curves (solid) of the properties of data set 1. The blue lines represent data set 6. (a) equivalent diameter; (b) the intensity contrast; (c) lifetime; (d) horizontal velocity; (e) diffusion index; (f) the ratio of motion range; (g) the index of motion type.}
        \label{Fig:fig4}
\end{figure}

Figure~\ref{Fig:fig4} only illustrates the histograms and the best fitted curves of data set 1 and 6 since their mean magnetic flux densities are the lowest and highest of all six data sets. The red lines represent the histograms (dashed) and the distribution curves (solid) of the properties of data set 1. The blue lines represent data set 6. The other data sets have the similar distributions and fitted curves.

Figure~\ref{Fig:fig4}(a) shows the distributions of equivalent diameters of igBPs and the log-normal curve fitted lines. The log-normal distribution of the igBP diameters is consistent with the suggestion that the process of fragmentation and merging dominate the process of flux concentration (\citealt{Abramenko2005Distribution}). The distribution of data set 1 is sharper than that of data set 6. The mean values and the standard deviations of data set 1 and 6 are 182$\pm$22 and 194$\pm$36\,$\rm km$, respectively. For all the six data sets, the mean diameters range from 168 to 195\,$\rm km$ (see Table~\ref{Tab:tab3}). All the minimum values saturate at scales corresponding to the diffraction limit of NVST, which is 105\,$\rm km$ in the TiO-band. The maximum value of data set 1 is 245\,$\rm km$, which is smaller than those of other data sets. We found that the isolated igBPs of data set 1 are nearly circular shape, while some of other data sets form chains that result in a relatively large equivalent diameter. These results are consistent with the previous studies (see Table~\ref{Tab:tab5}). All authors agreed that the minimum size of igBPs has not yet been detected in observations with modern high resolution telescopes and the maximum size cannot be larger than characteristic width of the intergranular lanes.

\begin{table}
\begin{center}
\caption[]{The Diameters of the IgBPs of Some Previous Studies}\label{Tab:tab5}
\begin{threeparttable}
  \begin{tabular}{lcccccc}
  \hline\noalign{\smallskip}
    Reference & Telescope & Region    &Magnetic & Spatial  & Temporal   & Mean diameter\\
            &           &           & fluxes        &resolution & resolution ($\rm s$)& ($\rm km$)\\
  \hline\noalign{\smallskip}
\citet{Berger1995New}           &SVST   &AR   &             & 0.$^{''}$083  &     &250\\
\citet{Bovelet2003Dynamics}     &DOT    &AR   &             & 0.$^{''}$071  &30    &220$\pm$25 \\
\citet{Feng2013Statistical}     &DOT    &QS    &            & 0.$^{''}$071  &30    &224$\pm$40  \\
\citet{Feng2013Statistical}     &DOT    &AR    &            & 0.$^{''}$071  &30    &232 $\pm$45 \\
\citet{Crockett2010Area}        &DST    &QS    &            & 0.$^{''}$069  &2      &230    \\
\citet{Utz2009size}             &SOT    &QS   &             & 0.$^{''}$054  &30    &166$\pm$31\\
\citet{Utz2009size}             &SOT    &QS   &             & 0.$^{''}$108  &30    &218$\pm$48\\
\citet{Romano2012Comparative}   &SST    &QS    &            & 0.$^{''}$041  &89    &216  \\
\citet{Romano2012Comparative}   &SST    &AR    &            & 0.$^{''}$041  &20    &268  \\
\citet{Abramenko2010Statistical} &NST   &QS    &            & 0.$^{''}$0375 &10    &77$-$260\tnote{1} \\
  \noalign{\smallskip}\hline
\end{tabular}
 \begin{tablenotes}
        \footnotesize
        \item[1] The range of igBP diameters.
 \end{tablenotes}
\end{threeparttable}
\end{center}
\end{table}

Figure~\ref{Fig:fig4}(b) shows the distributions of the intensity contrast of igBPs and the normal curve fitted lines. The distribution of data set 1 is sharper than that of data set 6. The mean values and the standard deviations of data set 1 and 6 are 0.99$\pm$0.04 and 1.06$\pm$0.05, respectively. It can be seen in Table~\ref{Tab:tab3} that the mean values of all the data sets range from 0.99 to 1.06, and the minimum value and the maximum value are 0.89 and 1.31, respectively. Previous studies concluded that the intensity contrast of G-band igBPs is 0.8-1.8 (see Table~\ref{Tab:tab6}). Both the minimum values are about 0.8, but the maximum value of TiO-band data is much less than 1.8. It implies that the intensity contrast of TiO-band igBPs is lower than G-band igBPs.

\begin{table}
\begin{center}
\caption[]{The Intensity contrasts of the IgBPs of Some Previous Studies}\label{Tab:tab6}
\begin{threeparttable}
  \begin{tabular}{lcccccc}
  \hline\noalign{\smallskip}
      Reference & Telescope & Region    &Magnetic & Spatial  & Temporal   &Mean intensity\\
            &           &           & fluxes        &resolution & resolution ($\rm s$)&contrast\\
              \hline\noalign{\smallskip}
      \citet{Feng2013Statistical}       &DOT   & QS    &        & 0.$^{''}$071  & 30    & 1.3 \\
      \citet{Feng2013Statistical}       &DOT   & AR    &        & 0.$^{''}$071  & 30    & 1.6 \\
      \citet{Utz2013Magnetic}           &SOT   & QS    &        & 0.$^{''}$108  & 30    & 1.0 \\
      \citet{Utz2013Magnetic}           &SOT   & AR    &        & 0.$^{''}$108  & 30    & 1.4 \\
      \citet{Yang2014Evolution}         &SOT   & QS    &        & 0.$^{''}$054  & 30    & 1.02$\pm$0.11\\
      \citet{Sanchez2004Bright}         &SST   & QS    & 0.9 G  & 0.$^{''}$041  & 15    & 0.8$-$1.8\tnote{1} \\
      \citet{Mostl2006Dynamics}         &SST   & AR    &        & 0.$^{''}$041  & 20    & 1.17$\pm$0.08 \\
      \citet{Romano2012Comparative}     &SST   & QS    &        & 0.$^{''}$041  & 20    & 1.09$\pm$0.05  \\
      \citet{Romano2012Comparative}     &SST   & AR    &        & 0.$^{''}$041  & 20    & 1.05$\pm$0.06  \\
  \hline\noalign{\smallskip}
\end{tabular}
 \begin{tablenotes}
        \footnotesize
        \item[1] The range of igBP intensity contrast.
 \end{tablenotes}
\end{threeparttable}
\end{center}
\end{table}

Figure~\ref{Fig:fig4}(c) shows the distributions of the lifetime of igBPs and the exponential curve fitted lines. The distributions indicate that most of igBPs have a short lifetime. The mean values of these similar distributions are 104 and 124$\,\rm s$, respectively. Note that, the standard deviation of an exponential function is as the same as the mean value. From Table~\ref{Tab:tab3}, it can be found that the mean lifetime values of all data sets range from 104 to 141\,$\rm s$, and the maximum value reaches 826\,$\rm s$. Our results are consistent with the most previous studies listed in Table~\ref{Tab:tab7}. However, some authors got a long lifetime value, e.g., \citet{Berger1998Measurements} and \citet{Nisenson2003Motions}. \citet{Berger1998Measurements} measured the mean lifetime for all igBPs including isolated and non-isolated igBPs. The lifetimes of non-isolated igBPs are generally long because they undergo numerous split or merge interactions. For the six data sets, the maximum lifetime of non-isolated igBPs is 59\,$\rm min$. \citet{Nisenson2003Motions} only measured the igBPs whose lifetimes are longer than 210\,$\rm s$. Here, we measured the lifetimes of isolated igBPs with lifetime $>$ 100\,$\rm s$.

\begin{table}
\begin{center}
\caption[]{The Lifetimes of the IgBPs of Some Previous Studies}\label{Tab:tab7}
\begin{threeparttable}
  \begin{tabular}{lcccccc}
  \hline\noalign{\smallskip}
        Reference & Telescope & Region    &Magnetic & Spatial  & Temporal   &Mean lifetime\\
            &           &           & fluxes        &resolution & resolution ($\rm s$)&($\rm s$)\\
    \hline\noalign{\smallskip}
\citet{Berger1998Measurements}      &SVST       &QS     &           &0.$^{''}$083   &23     &560 \\
\citet{Nisenson2003Motions}         &DOT        &Network     &      &0.$^{''}$071   &30     &552 \\
\citet{deWijn2005DOT}               &DOT        &Network     &      &0.$^{''}$071   &30     &210 \\
\citet{Keys2014Dynamic}             &DST        &QS     &3 G        &0.$^{''}$069   &2      &88$\pm$23  \\
\citet{Keys2014Dynamic}             &DST        &AR     &169 G      &0.$^{''}$069   &2      &136$\pm$40 \\
\citet{Utz2010Dynamics}             &SOT        &QS     &           &0.$^{''}$054   &30     &150$\pm$5 \\
\citet{Mostl2006Dynamics}           &SST        &AR    &            &0.$^{''}$041   &20     &260$\pm$137 \\
\citet{Abramenko2010Statistical}    &NST        &QS     &           &0.$^{''}$0375  &10     &120$-$720\tnote{1} \\
  \noalign{\smallskip}\hline
\end{tabular}
\begin{tablenotes}
        \footnotesize
        \item[1] The range of igBP lifetime.
 \end{tablenotes}
\end{threeparttable}
\end{center}
\end{table}

Figure~\ref{Fig:fig4}(d) shows the distributions of all the calculated horizontal velocities between successive frames of igBPs. The horizontal velocities in the $x$ and the $y$ direction are both fitted to a normal function well. Therefore, the distribution of rms horizontal velocities fits the Rayleigh function well. The distribution of data set 6 is sharper than that of data set 1, which is different from equivalent diameters and the intensity contrast. The mean values and the standard deviations of data set 1 and 6 are 1.35$\pm$0.71 and 1.05$\pm$0.55\,$\rm km s^{-1}$, respectively. For all the data sets, the mean values range from 1.04 to 1.35\,$\rm km s^{-1}$, which confirm the previous studies(see Table~\ref{Tab:tab8}).

\begin{table}
\begin{center}
\caption[]{The horizontal velocities of the IgBPs of Some Previous studies}\label{Tab:tab8}
  \begin{tabular}{lcccccc}
  \hline\noalign{\smallskip}
          Reference & Telescope & Region    &Magnetic & Spatial  & Temporal   &Mean horizontal velocity\\
            &           &           & fluxes        &resolution & resolution ($\rm s$)&($\rm km$ $\rm s^{-1}$)\\
  \hline\noalign{\smallskip}
\citet{Berger1998Measurements}          &SVST     &QS     &       &0.$^{''}$083   &23     &1.1 \\
 \citet{Berger1998Measurements}         &SVST     &AR     &       &0.$^{''}$083   &23     &0.95 \\
 \citet{Nisenson2003Motions}            &DOT      &Network     &  &0.$^{''}$071   &30     &0.89 \\
 \citet{Keys2014Dynamic}                &DST      &QS     &3 G    &0.$^{''}$069    &2      &0.9$\pm$0.4 \\
\citet{Keys2014Dynamic}                 &DST      &AR     &169 G  &0.$^{''}$069   &2      &0.6$\pm$0.3   \\
\citet{Utz2010Dynamics}                 &SOT      &QS    &        &0.$^{''}$054   &30     &1.62$\pm$0.05\\
\citet{Mostl2006Dynamics}               &SST      &AR    &        &0.$^{''}$041   &20     &1.11 \\
  \noalign{\smallskip}\hline
\end{tabular}
\end{center}
\end{table}

Figure~\ref{Fig:fig4}(e) shows the distributions of the diffusion index, $\gamma$, and the normal curve fitted lines. Diffusion processes represents the efficiency of dispersal in the photosphere, which uses a diffusion index to quantify the transport process with respect to a normal diffusion (random walk). Diffusion process can be characterized by the relation $(\Delta l)^{2} = C\tau^{\gamma}$, where $\Delta l$ represents the displacement of an igBP between its location at given time $\tau$ and its initial location; $\gamma$ is the diffusion index and $C$ is a constant of proportionality (\citealt{Dybiec2009Discriminating}; \citealt{Jafarzadeh2014Migration}; \citealt{Yang2015Dispersal}). Diffusions with $\gamma <$ 1, $\gamma =$ 1 and $\gamma >$ 1 are called sub-diffusive, normal-diffusive and super-diffusive, respectively. The $\gamma$ value of each igBP was measured by the slope of its square displacement on a log-log scale, where the time scale is determined by its lifetime. Subsequently, we obtained the mean $\gamma$ value and the standard deviation by curve fitting the histogram of all $\gamma$ values. The distributions are fitted to a normal function well. As a result, the mean $\gamma$ values of data set 1 and 6 are 1.31$\pm$0.65 and 0.93$\pm$0.77, respectively. It also can be seen that the mean values of the six data sets range from 0.86 to 1.31 in Table~\ref{Tab:tab3}. It is worth noting that the $\gamma$ values with less 0 are meaningless. If an igBP moves in an erratic or circular path, the slope of linear fit on a log-log scale would be below 0 or very large with a low goodness-of-fit (\citealp{Yang2015Dispersal}). The results are in agreement with the most previous studies (see Table~\ref{Tab:tab9}). The authors got the range of diffusion index from 0.76 to 1.79. It implies that igBPs in different magnetic regions have various regimes, such as sub-, normal- and super-diffusive regime. IgBPs in strong magnetic fields are crowded within narrow intergranular lanes when compared with a weak magnetic environment, where igBPs can move freely due to a lower population density (\citealt{Abramenko2011Turbulent}). It is the main reason that igBPs in a weaker magnetic region diffuse faster than that in a stronger one.

\begin{table}
\begin{center}
\caption[]{The diffusion indices of the IgBPs of Some Previous Studies}\label{Tab:tab9}
\begin{threeparttable}
  \begin{tabular}{lcccccc}
  \hline\noalign{\smallskip}
            Reference & Telescope & Region    &Magnetic & Spatial  & Temporal   &Mean diffusion\\
            &           &           & fluxes        &resolution & resolution ($\rm s$)&index\\
  \hline\noalign{\smallskip}
\citet{Cadavid1999Anomalous}      &SVST     &Network     &      &0.$^{''}$083   &23     &0.76$\pm$0.04$/$1.10$\pm$0.24\tnote{1} \\
\citet{Keys2014Dynamic}           &DST      &QS     &3 G        &0.$^{''}$069   &2      &1.21$\pm$0.25   \\
\citet{Keys2014Dynamic}           &DST      &AR     &169 G      &0.$^{''}$069   &2      &1.23$\pm$0.22   \\
\citet{Yang2015Dispersal}         &SOT      &QS     &           &0.$^{''}$108   &30     &1.79$\pm$0.01\\
\citet{Yang2015Dispersal}         &SOT      &AR     &           &0.$^{''}$108   &30     &1.53$\pm$0.01\\
\citet{Abramenko2011Turbulent}    &NST      &coronal hole &     &0.$^{''}$0375   &10     &1.67 \\
\citet{Abramenko2011Turbulent}    &NST      &QS     &           &0.$^{''}$0375   &10     &1.53 \\
\citet{Abramenko2011Turbulent}    &NST      &plage area     &   &0.$^{''}$0375   &10     &1.48 \\
  \noalign{\smallskip}\hline
\end{tabular}
\begin{tablenotes}
        \footnotesize
        \item[1] for time intervals of 0.3$-$22 minutes / for 25$-$57 minutes.
 \end{tablenotes}
\end{threeparttable}
\end{center}
\end{table}

Figure~\ref{Fig:fig4}(f) shows the distributions of the ratio of motion range of igBPs and the log-normal curve fitted lines.
The rate of motion range is defined as $m_{r} =\sqrt{(X_{max}-X_{min})^{2}+(Y_{max}-Y_{min})^{2}}/r$, where $X_{max}$ and $X_{min}$ are the maximum and minimum coordinates of the path of a single igBP in the $\bm{x}$ axis, and $Y_{max}$ and $Y_{min}$ are in the $\bm{y}$ axis; $r$ is the radius of the circle which corresponds to the maximum size of the igBP during its lifetime (\citealt{Bodnarova2014On}; \citealt{Yang2015Characterizing}). If the $m_{r}$ value of an igBP is less than 1, the igBP moves within its own maximum radius during its lifetime. Such igBP is called stationary one. Otherwise, igBP with $m_{r} \geq$ 1 is called non-stationary one. The numbers of non-stationary igBPs of the six data sets are listed in Table~\ref{Tab:tab2}. The percents of the non-stationary igBPs range from 40$\%$ to 73$\%$. \citet{Bodnarova2014On} and \citet{Yang2015Characterizing} both found that the $m_{r}$ values of about 50\% igBPs are less than 1 from QS observations. We believed that our wide-range results are caused by the different magnetic environments. Moreover, the mean $m_{r}$ values and the standard deviations of data set 1 and data set 6 are 1.30$\pm$0.80 and 1.03$\pm$0.69, and the maximum $m_{r}$ values are 5.06 and 4.79, respectively. From Table~\ref{Tab:tab3}, the mean values of the six data sets range from 0.96 to 1.30, which implies that igBPs move nearly as far as its radius. The maximum values of the six data sets range from 4.73 to 6.39, which are consistent with the previous studies that got a value of about 7 (\citealt{Bodnarova2014On}; \citealt{Yang2015Characterizing}).

Figure~\ref{Fig:fig4}(g) shows the distributions of the index of motion type of igBPs and the exponential curve fitted lines.
Putting aside the stationary igBPs, we focused on the motion type of the non-stationary ones. The index of motion type is defined as $m_{t}=d/L$, where $d=\sqrt{(X_{n}-X_{1})^{2}+(Y_{n}-Y_{1})^{2}}$, here $(X_{1},Y_{1})$ is the start location and $(X_{n},Y_{n})$ is the final location during its lifetime; $L$ is the whole path length, defined as $L=\sum_{t=1}^{n-1}\sqrt{\triangle X_{t}^{2}+\triangle Y_{t}^{2}}$, here $\triangle X_{t}=X_{t+1}-X_{t}$, $\triangle Y_{t}=Y_{t+1}-Y_{t}$. It is a ratio of the displacement of an igBP to its whole path length \citep{Yang2015Characterizing}. According to the definition, the $m_{t}$ value must be between 0 and 1. If an igBP moves in a nearly straight line, the $m_{t}$ value will be close to 1. But if it moves in a nearly closed curve, then the $m_{t}$ value will be close to 0. The motion types of non-straight lines refer to as erratic motion type. Please refer to \citet{Yang2015Characterizing} for details. We found that the distributions of $m_{t}$ values fit to exponential function well. The $m_{t}$ values of half of igBPs of data set 1 and 6 are greater than 0.75 and 0.68, respectively. About 25$\%$ and 29$\%$ igBPs of data set 1 and 6 have the $m_{t}$ values less than 0.5, respectively. The mean values are 0.69 and 0.62, respectively. The mean $m_{t}$ values of the six data sets range from 0.58 to 0.69. The distribution of $m_{t}$ is very similar to that analyzed by \citet{Yang2015Characterizing}. They indicated that the $m_{t}$ value of half igBPs are larger than 0.83 and 15$\%$ are less than 0.5.

We applied different curve fitted functions to determine the analytical fit for the distribution functions of the properties of igBPs, such as normal, log-normal, exponential, and Rayleigh. The parameter of Adjust R-square, which is called the adjusted square of the multiple correlation coefficient, indicates how successful the fit is in explaining the variation of the data (\citealt{cameron1997r}). Consequently, we adopted the functions with the highest Adjust R-square values: normal function for intensity contrasts and diffusion indices, log-normal for equivalent diameters and the ratios of motion range, exponential for lifetimes and the indices of motion type, and Rayleigh for horizontal velocities. These distribution functions are consistent with the most previous works cited above.

\subsection{Relation between igBP properties and magnetic field}
\label{sect:relation}
\begin{figure}[!htp]
        \centering
        \includegraphics[width=0.4\textwidth]{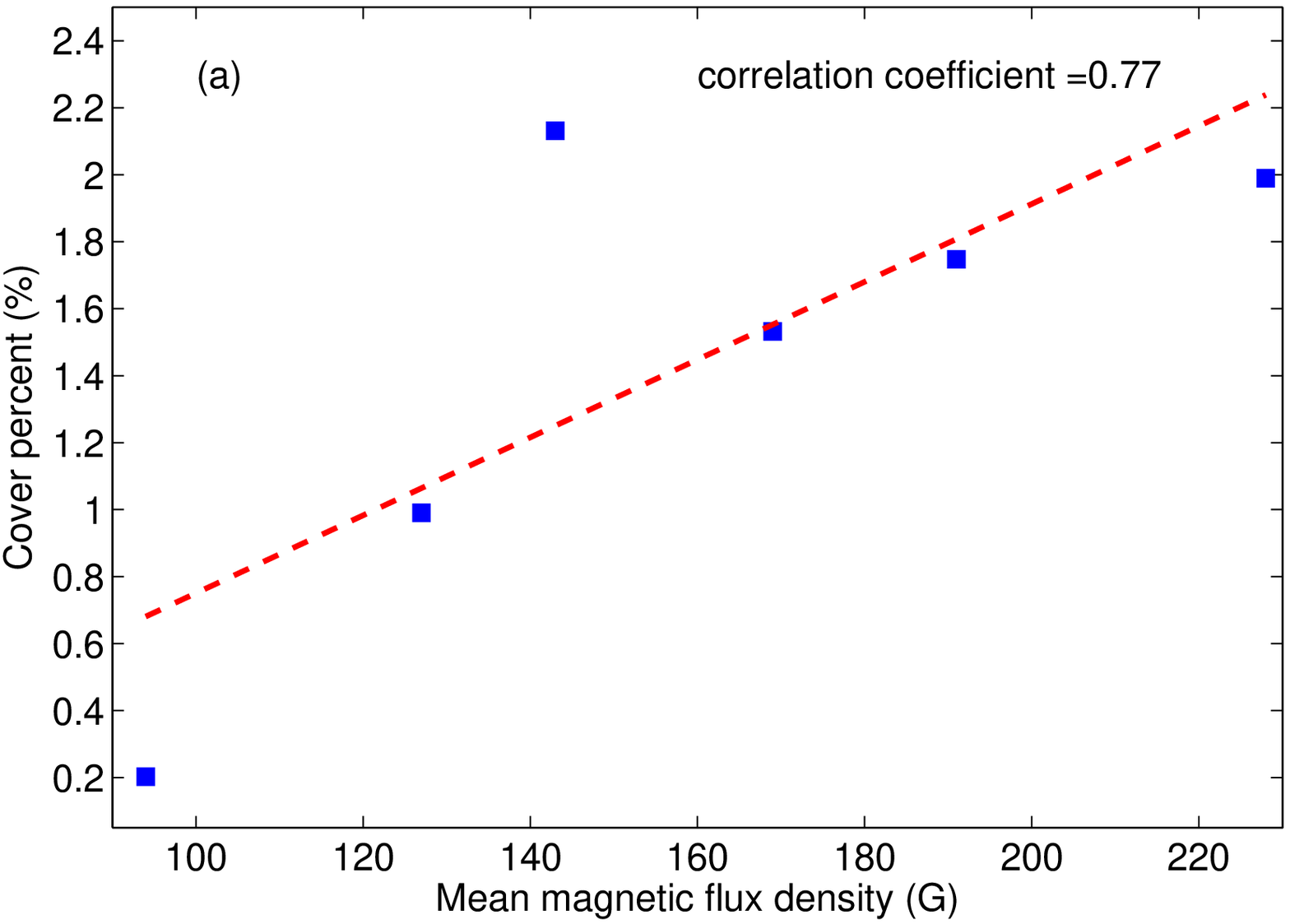}
        \includegraphics[width=0.4\textwidth]{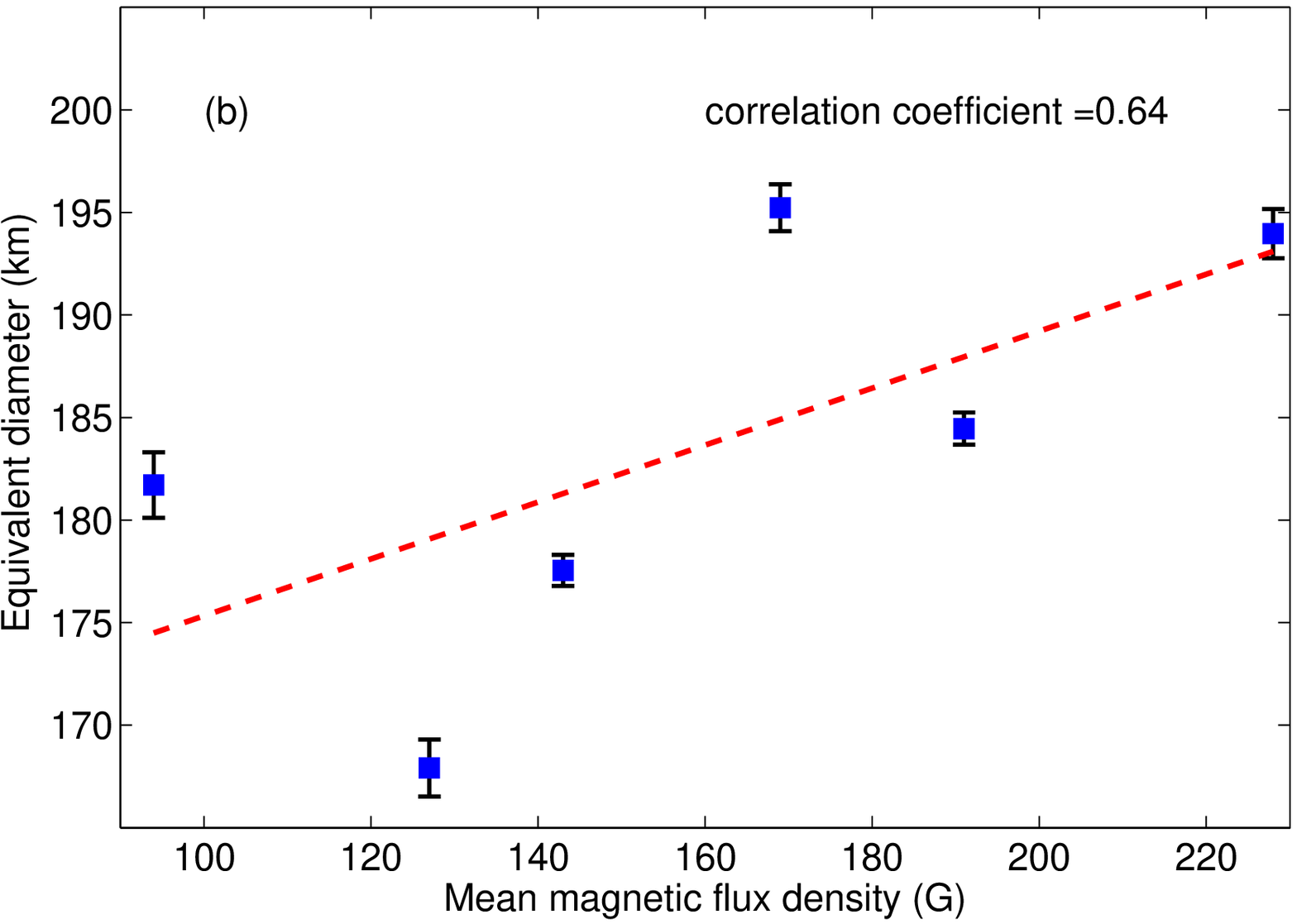}
        \includegraphics[width=0.4\textwidth]{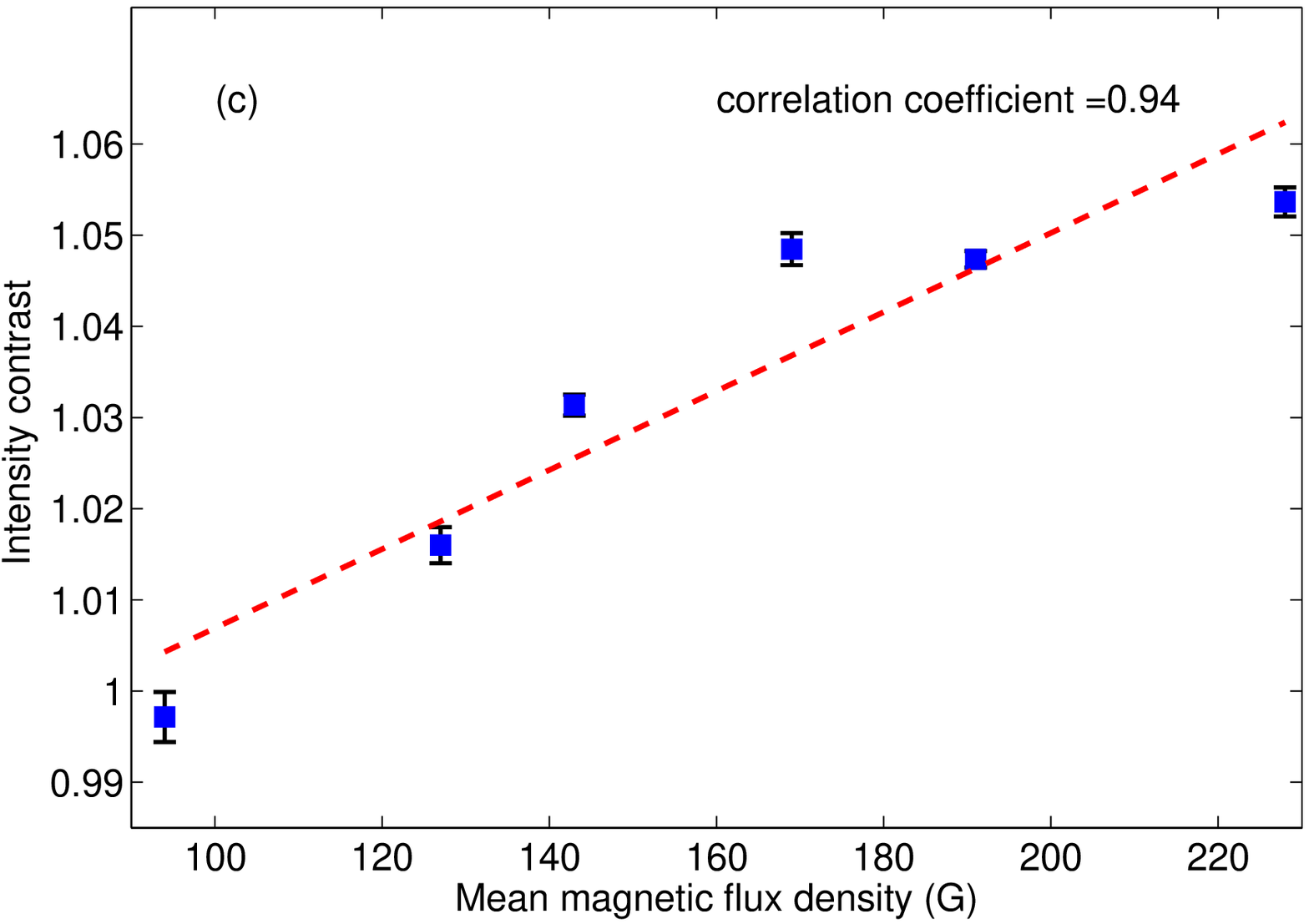}
        \includegraphics[width=0.4\textwidth]{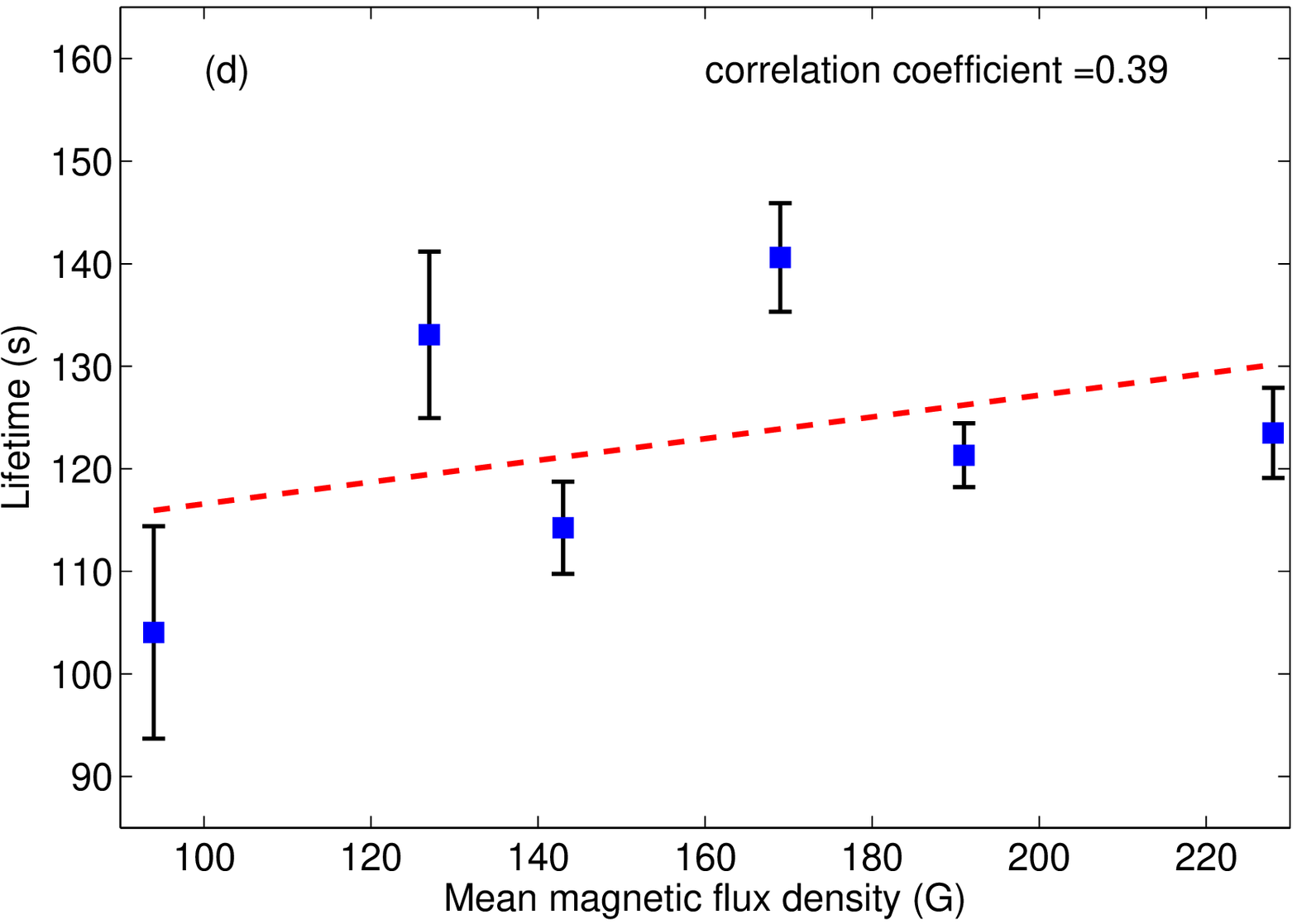}
        \includegraphics[width=0.4\textwidth]{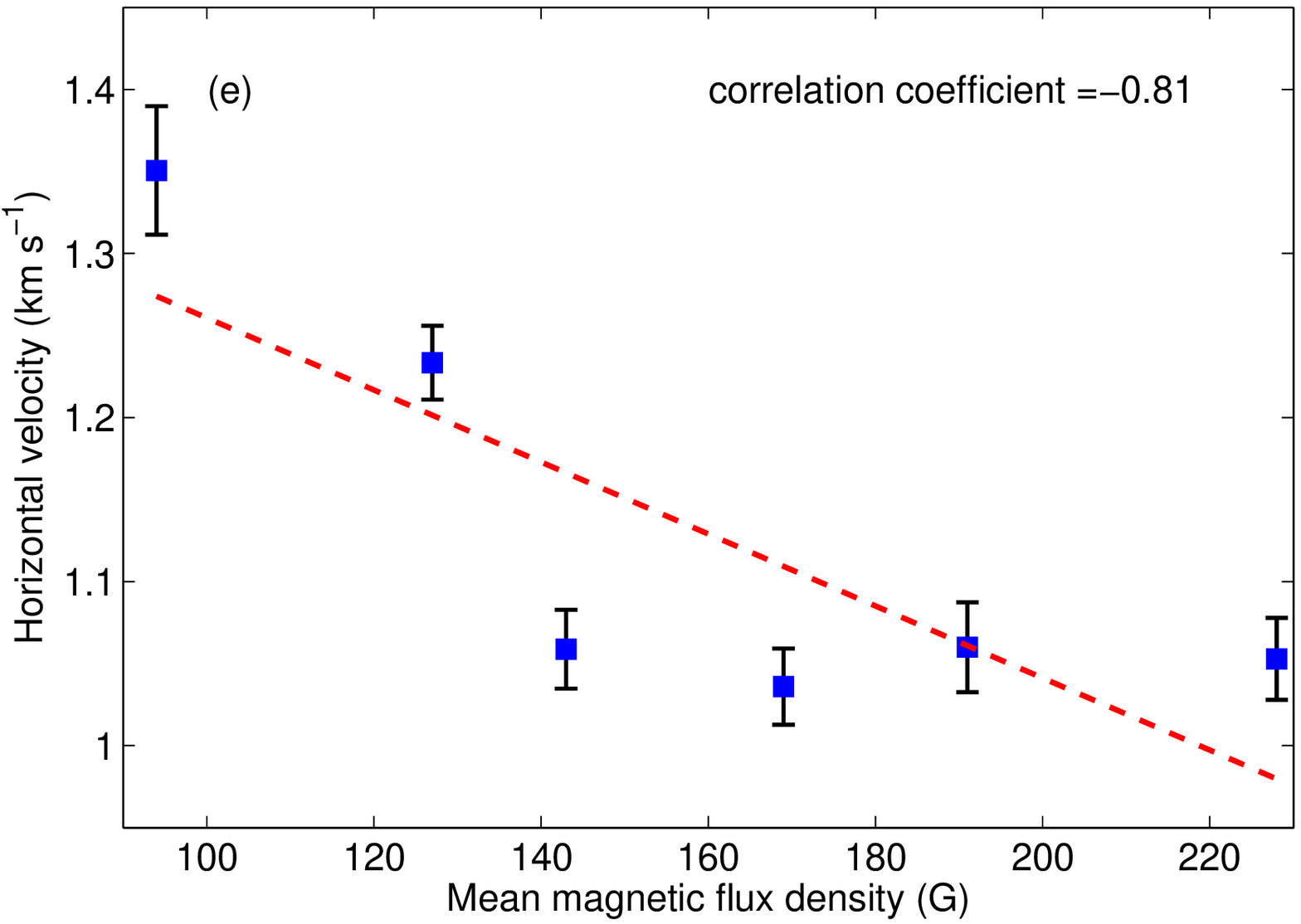}
        \includegraphics[width=0.4\textwidth]{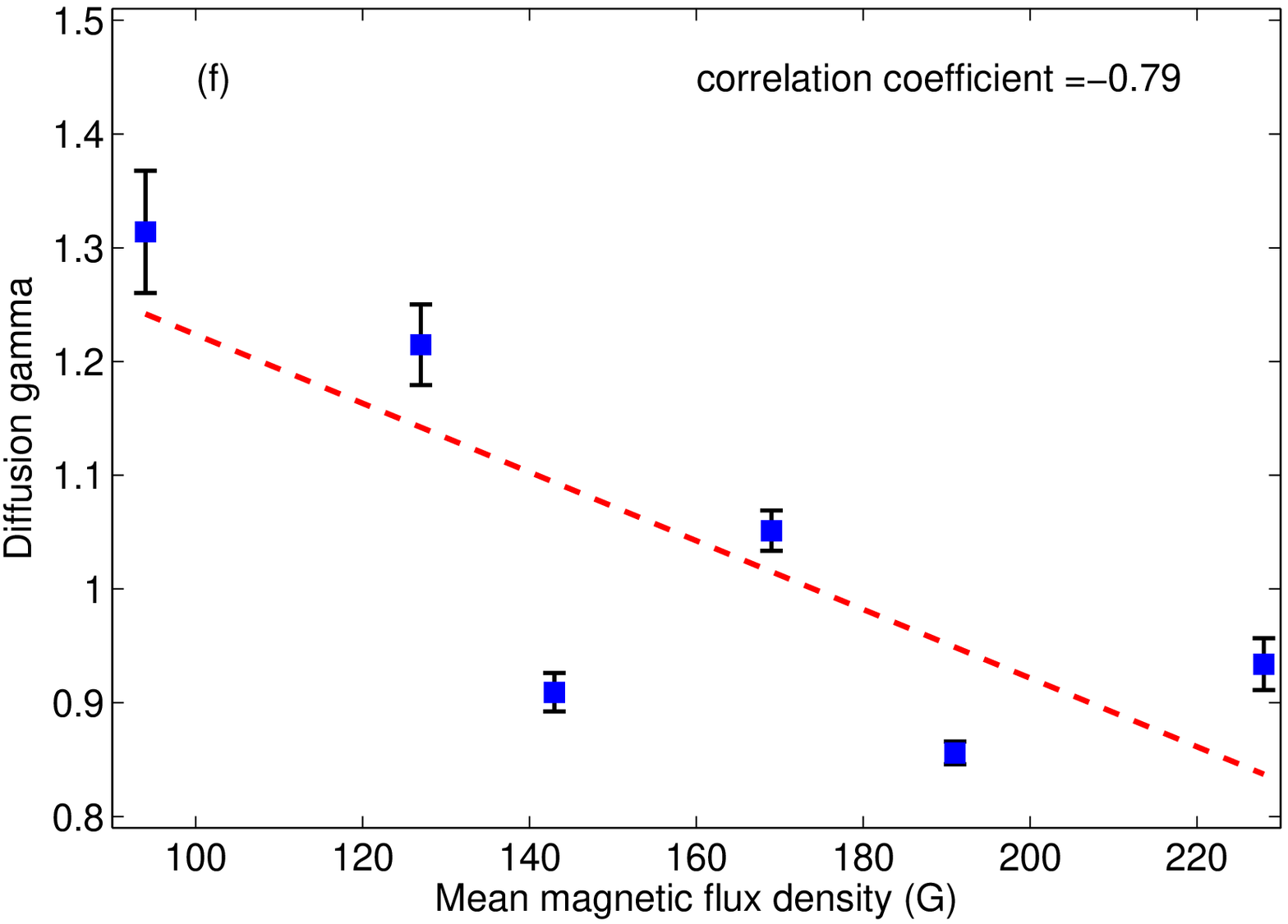}
        \includegraphics[width=0.4\textwidth]{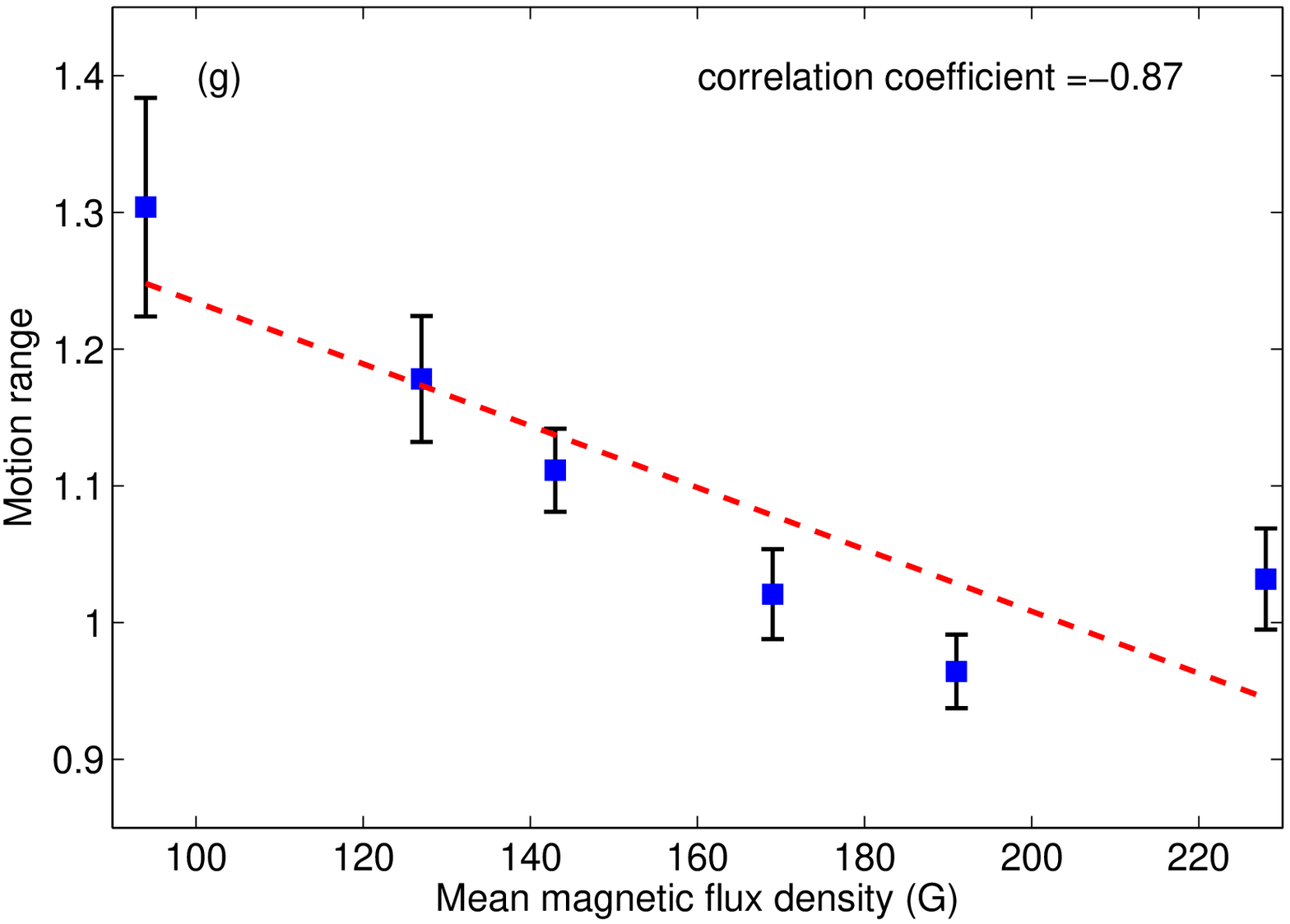}
        \includegraphics[width=0.4\textwidth]{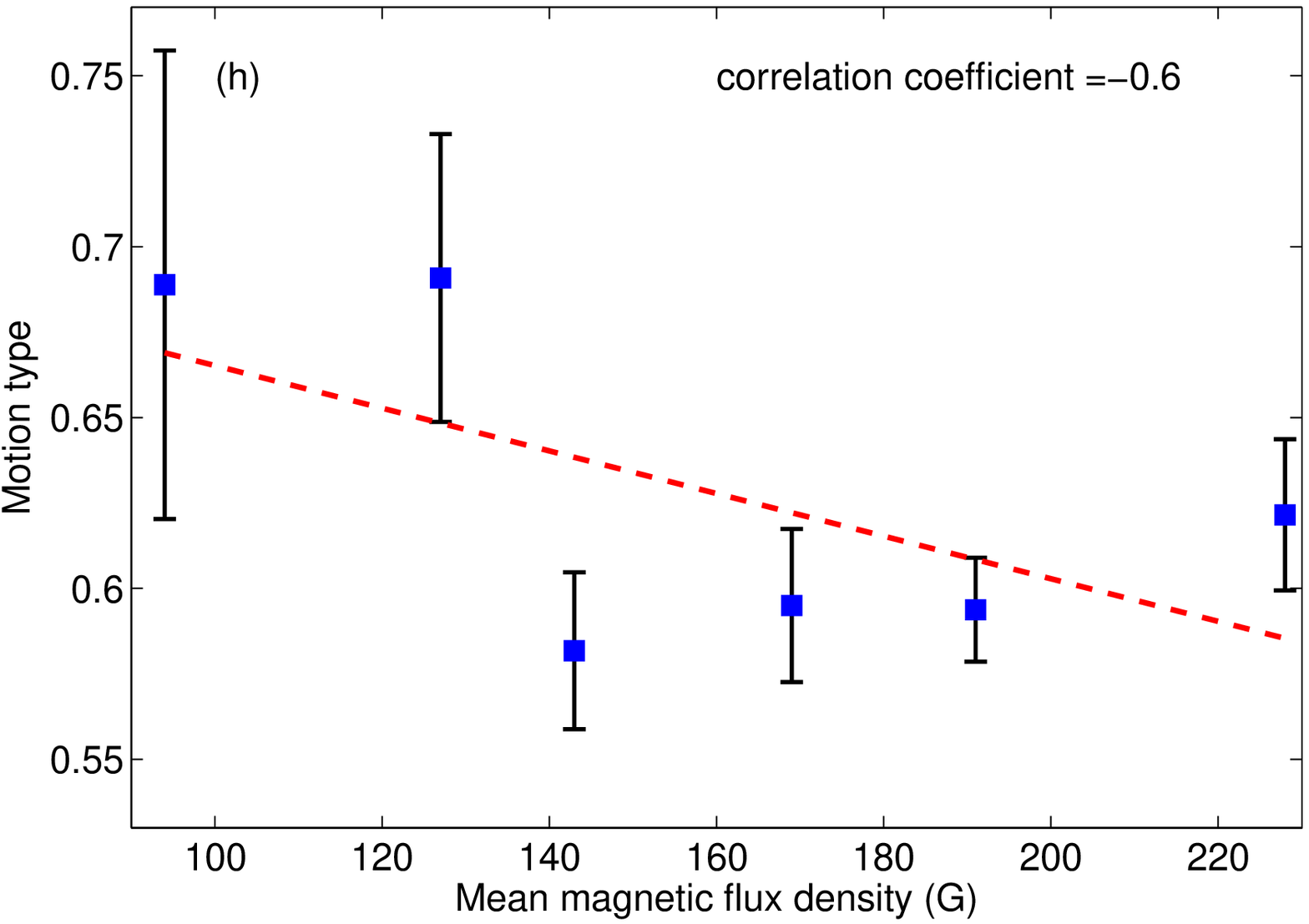}
        \caption{The relation plots between the mean magnetic flux density and the igBP properties: (a) area coverage-mean magnetic flux density; (b) diameter-mean magnetic flux density; (c) intensity contrast-mean magnetic flux density; (d) lifetime-mean magnetic flux density; (e) horizontal velocity-mean magnetic flux density; (f) diffusion gamma-mean magnetic flux density; (g) the ratio of motion range-mean magnetic flux density; (h) the index of motion type-mean magnetic flux density. The blue box and the black line are the mean value and the standard error of igBP property of each data set, respectively. The solid red line is the linear fitted one. }
        \label{Fig:fig5}
\end{figure}

It can be seen that the statistical values of igBPs in the six magnetic environments are different in Table~\ref{Tab:tab3} and Figure~\ref{Fig:fig4}. In order to explore the relations between the properties of igBPs and their embedded magnetic environments, Figure~\ref{Fig:fig5} shows the correlations between the mean magnetic flux density of the region that igBPs are embedded and igBP properties of the six data sets, in terms of area coverage, diameter, intensity contrast, lifetime, horizontal velocity, diffusion index, the ratio of motion range and the index of motion type. The blue box and the black line are the mean value and the standard error of each data set, respectively. The solid red line is the linear fitted one. The relations of area coverage-magnetic flux density, diameter-magnetic flux density, the intensity contrast-magnetic flux density show a positive correlation with the correlation coefficients of 0.77, 0.64 and 0.94, respectively. The relations of horizontal velocity-magnetic flux density, diffusion index-magnetic flux density, the ratio of motion range-magnetic flux density and the index of motion type-magnetic flux density show a negative correlation with the correlation coefficients of -0.81, -0.79, -0.87 and -0.60, respectively. However, the lifetime-magnetic flux density relation fail to exhibit an obvious correlation with the correlation coefficient of 0.39.

Our results are in good qualitative agreement with the most previous studies in Table~\ref{Tab:tab4}$-$Table~\ref{Tab:tab9}.
It implies that in a higher magnetic environment, the igBP is larger and brighter, and its movement is attenuated (e.g. the lower horizontal velocity, the sub-diffusion, the limited motion range and the erratic motion type). These different physical properties result from the inhibition of convection induced by the presence of the magnetic field, which changes the temperature stratification of both quiet and magnetic regions (\citealt{Criscuoli2013Comparison}). Note that, we found that the intensity contrast of igBPs in a higher magnetic environment is larger than that in a lower magnetic environment. The result is consistent with \citet{Feng2013Statistical} and \citet{Utz2013Magnetic}, but differs from \citet{Romano2012Comparative}. Additionally, the previous studies suggested that the igBPs live longer in stronger magnetic environments (\citealt{Keys2014Dynamic}), however, our results did not show an obvious relation on the embedded magnetic environment.

\section{Conclusion}
\label{sect:conclusion}

Six high-resolution TiO-band image sequences which span from 2012 to 2014 year were obtained under excellent seeing conditions from the New Vacuum Solar Telescope (NVST) in Fuxian Solar Observatory of Yunnan Astronomical Observatory, China. We investigate the morphologic, photometric and dynamic properties of igBPs, in terms of equivalent diameter, the intensity contrast, lifetime, horizontal velocity, diffusion index, motion range and motion type. With the aid of vector magnetograms obtained with the SDO / HMI, the statistical properties of igBPs in different magnetic environments are also explored and compared.

The statistics of igBPs indicate that the quality of the TiO-band data from the NVST is stable and reliable. The area coverages of igBPs range from 0.2\% to 2\%. The mean equivalent diameters range from 168$\pm$29 to 195$\pm$36\,$\rm km$. The mean ratios of intensity contrast range from 0.99$\pm$0.04 to 1.06$\pm$0.05. The mean lifetimes range from 104 to 141\,$\rm s$. The mean horizontal velocities range from 1.04$\pm$0.54 to 1.35$\pm$0.71\,$\rm km s^{-1}$. The mean diffusion indices range from 0.86$\pm$0.39 to 1.31$\pm$0.54. The mean ratio of motion range values range from 0.96$\pm$0.67 to 1.30$\pm$0.80, and the mean index of motion type values range from 0.58 to 0.69. Moreover, the mean values and the standard deviations of igBP properties are consistent with the previously published studies based on G-band or TiO-band observations from the other telescopes, such as SVST, DOT, DST, SOT, SST, NST. It implies that the TiO-band data from the NVST are very suitable to study the igBPs, and the LMD and three-dimensional segmentation algorithms are feasible to detect and track the igBPs from the TiO-band data from the NVST.

In addition, different magnetic environments are considered, characterized by different mean magnetic flux density. The area coverage, the size and the intensity contrast values of igBPs are generally larger in the regions of high magnetic flux. However, the dynamics of igBPs, in terms of the horizontal velocity, the diffusion index, the ratio of motion range and the index of motion type are generally larger in the regions of low magnetic flux.
Previous studies focused on comparing the properties of igBPs in the quiet Sun and active region, or different sub-regions in the same FOV. This study provides further information about the relation between the properties of igBPs and their embedded magnetic environments based on six data sets that span three years, located in different solar positions and have different magnetic fluxes. It suggests that the stronger magnetic field makes the igBPs look bigger and brighter, attenuate their movements (e.g. the lower horizontal velocity, the sub-diffusion, the limited motion range and the erratic motion type).

\begin{acknowledgements}
The authors are grateful to the anonymous referee for constructive comments and detailed suggestions to this manuscript. The authors are grateful to the support received from the National Natural Science Foundation of China (No: 11573012, 11303011, 11263004, 11163004, U1231205), Open Research Program of the Key Laboratory of Solar Activity of the Chinese Academy of Sciences (No: KLSA201414, KLSA201505). The authors thank the NVST team for their high-resolution observations and level 1$^{+}$ data. The HMI data used here are courtesy of NASA/SDO and the HMI science teams.
\end{acknowledgements}


\label{lastpage}

\end{document}